\documentclass[journal,a4paper,onecolumn,11pt]{IEEEtran}

\usepackage[cmex10]{amsmath} 
\allowdisplaybreaks
\usepackage{amssymb,mathtools}
\usepackage{mathrsfs}     
\usepackage[T1]{fontenc}  
\usepackage{bm}           
\usepackage{bbm}          
\usepackage{comment}      
\usepackage[noadjust]{cite} 
\usepackage{url}
\usepackage{hyperref}
\hypersetup{colorlinks=true,linkcolor=blue,filecolor=blue,citecolor=blue,urlcolor=black}

\usepackage{ifthen}
\usepackage{theorem}
\theoremstyle{plain}
\newtheorem{definition}{Definition}
\newtheorem{theorem}[definition]{Theorem}
\newtheorem{proposition}[definition]{Proposition}
\newtheorem{lemma}[definition]{Lemma}

\newtheorem{corollary}[definition]{Corollary}

\newtheorem{remark}[definition]{Remark}

\mathchardef\ordinarycolon\mathcode`\:
\mathcode`\:=\string"8000
\def\vcentcolon{\mathrel{\mathop\ordinarycolon}}
\begingroup \catcode`\:=\active
  \lowercase{\endgroup
  \let :\vcentcolon
}

\newcommand{\Renyi}{R\'{e}nyi~}
\newcommand{\wt}[1]{\widetilde{#1}}
\newcommand{\wh}[1]{\widehat{#1}}
\newcommand{\ol}[1]{\overline{#1}}
\newcommand{\etal}{et al.}
\newcommand{\ket}[1]{\left\vert #1 \right\rangle}
\newcommand{\bra}[1]{\left\langle #1 \right\vert}

\newcommand\proj[1]{\vert #1 \rangle\!\langle #1 \vert}
\newcommand{\linear}[1]{\mathscr{L}(#1)}

\newcommand{\pos}[1]{\mathscr{P}(#1)}
\newcommand{\density}[1]{\mathscr{D}(#1)}
\newcommand{\subdensity}[1]{\mathscr{D}_{\leq}(#1)}  
\newcommand{\ox}{\otimes}
\newcommand{\1}{\mathbbm{1}}
\DeclareMathOperator{\tr}{Tr}  
\newcommand{\id}{\operatorname{id}}
\newcommand{\rel}{\middle\|}
\newcommand{\SEP}{\operatorname{SEP}} 

\DeclareMathOperator{\Shannon}{H}
\DeclareMathOperator{\Binary}{H_{bin}}
\DeclareMathOperator{\Mutual}{I} 
\DeclareMathOperator{\Rel}{D}       
\newcommand{\SRRel}[1]{\ensuremath{\widetilde{\operatorname{D}}_{#1}}}

\newcommand*{\cD}{\mathcal{D}}
\newcommand*{\cE}{\mathcal{E}}
\newcommand*{\cH}{\mathcal{H}}
\newcommand*{\cM}{\mathcal{M}}
\newcommand*{\cN}{\mathcal{N}}
\newcommand*{\cP}{\mathcal{P}}
\newcommand*{\cT}{\mathcal{T}}
\newcommand*{\cW}{\mathcal{W}}
\newcommand*{\cX}{\mathcal{X}}
\newcommand*{\cZ}{\mathcal{Z}}

\newcommand*{\bRp}{\mathbb{R}_{+}}
\newcommand*{\bNp}{\mathbb{N}_{+}}

\begin{document}
\title{Permutation Enhances Classical Communication Assisted by Entangled States}

\author{%
  \IEEEauthorblockN{
    Kun Wang\IEEEauthorrefmark{1}\IEEEauthorrefmark{2} and
    Masahito Hayashi\IEEEauthorrefmark{3}\IEEEauthorrefmark{1}\IEEEauthorrefmark{4}\IEEEauthorrefmark{2}}

  \IEEEauthorblockA{\IEEEauthorrefmark{1}%
    Shenzhen Institute for Quantum Science and Engineering,\\
    Southern University of Science and Technology, Shenzhen 518055, China}

  \IEEEauthorblockA{\IEEEauthorrefmark{2}%
    Center for Quantum Computing, Peng Cheng Laboratory, Shenzhen 518055, China}

\IEEEauthorblockA{\IEEEauthorrefmark{3}%
    Graduate School of Mathematics, Nagoya University, Nagoya, 464-8602, Japan}

\IEEEauthorblockA{\IEEEauthorrefmark{4}%
    Centre for Quantum Technologies, National University of Singapore, \\
    3 Science Drive 2, 117542, Singapore}

\url{masahito@math.nagoya-u.ac.jp}
}

\maketitle

\begin{abstract}
We give a capacity formula for the classical communication over a noisy quantum channel, when local operations and
global permutations allowed in the encoding and bipartite states preshared between the sender and the receiver. The two
endpoints of this formula are the Holevo capacity (without entanglement assistance) and the entanglement-assisted
capacity (with unlimited entanglement assistance). What's more, we show that the capacity satisfies the strong converse
property and thus the formula serves as a sharp dividing line between achievable and unachievable rates of
communication. We prove that the difference between the assisted capacity and the Holevo capacity is upper bounded by
the discord of formation of the preshared state. As examples, we derive analytically the classical capacity of various
quantum channels of interests. Our result witnesses the power of random permutation in classical communication,
whenever entanglement assistance is available.
\end{abstract}


\section{Introduction}\label{sec:introduction}

\IEEEPARstart{A}{fundamental} task in information theory is to characterize the capability of transmitting classical
message over a channel in the asymptotic limit. The Shannon's noisy channel coding
theorem~\cite{shannon1948mathematical,fano1952class,wolfowitz1957coding} stated that the capacity of a classical
channel is representable as a single-letter quantity, capturing the amount of message that can be transmitted. Quantum
channels, however, do not have a single quantity characterizing their capacity for classical information transmission.
The Holevo-Schumacher-Westmoreland (HSW) theorem~\cite{schumacher1997sending,holevo1998capacity} established that the
classical capacity of a quantum channel is given by the regularized Holevo information of the channel.

When entanglement comes to play, the classical communication over quantum channels becomes much more profound. Unlike
shared randomness cannot increase a classical channel's capacity~\cite{shannon1948mathematical}, shared entanglement
will generally increase the classical communication rate of a quantum channel. For example, the superdense
coding~\cite{bennett1992communication} reveals that two classical bits can be sent through a single use of a noiseless
qubit channel, when assisted by a Bell state. The Bennett-Shor-Smolin-Thapliyal (BSST)
theorem~\cite{bennett1999entanglement,bennett2002entanglement} established a single-letter formula quantifying the
capacity of a quantum channel for classical communication, under the assumption that \textit{unlimited} entanglement
assistance is available. 

On the other hand, less is known on the limited entanglement-assisted classical communication regime: if the preshared
entanglement between the sender and the receiver is limited, or even noisy, how can we make use of this entanglement
assistance and how much classical information can be transmitted? Shor~\cite{shor2004classical} gave a trade-off curve
showing the classical capacity of a quantum channel as a function of the amount of available entanglement preshared.
The entanglement is measured in ebits. Furthermore, Zhu \etal~\cite{zhu2017superadditivity} constructed a channel for
which the classical capacity is additive, but that with limited entanglement assistance can be superadditive. Zhuang
\etal~\cite{zhuang2017additive} gave an additive capacity formula for the classical communication, with separable
encoding by the sender and limited resources supplied by the receiver's preshared ancilla. B\"{a}uml
\etal~\cite{bauml2019every} showed that for any entangled state, one can always construct a quantum memory channel
whose the feedback-assisted classical capacity can be increased by using the state as assistance.

In this work we push forward the study of limited entanglement-assisted classical communication by deriving a capacity
formula for entangled state assisted classical communication over a noisy quantum channel, when the encoding operations
are restricted to local operations and global permutations and multiple copies of an entangled state are preshared
among the sender and the receiver in product form. The two endpoints of this formula are the Holevo capacity -
corresponding to the case without entanglement assistance - and the entanglement-assisted capacity - corresponding to
the case with unlimited entanglement assistance. Our result reveals that whenever entanglement assistance is available,
global permutation can enhance classical communication compared to the case when only local encoding is allowed. What's
more, we show that the capacity satisfies the strong converse property and thus the formula serves as a sharp dividing
line between achievable and unachievable rates of communication. We also quantitatively investigate the gap between the
assisted capacity and the Holevo capacity, aiming to explore the preshared state's ability in enhancing classical
communication. We prove that this gap is upper bounded by the discord of formation of the preshared state.

\paragraph*{Notation} For a finite-dimensional Hilbert space $\cH$, we denote by $\linear{\cH}$ and $\pos{\cH}$ the
linear and positive semidefinite operators on $\cH$. Quantum states are in the set
$\density{\cH}:=\{\rho\in\pos{\cH}\vert\tr\rho=1\}$ and we also define the set of subnormalized quantum states
$\subdensity{\cH}:=\{\rho\in\pos{\cH}\vert0<\tr\rho\leq1\}$. For two operators $A, B\in\linear{\cH}$, we write $A\geq
B$ if and only if $A-B\in\pos{\cH}$. The identity matrix is denoted as $\1$ and the maximally mixed state is denoted as
$\pi$. Multipartite quantum systems are described by tensor product spaces. We use capital letters to denote the
different systems and subscripts to indicate on what subspace an operator acts. For example, if $L_{AB}$ is an operator
on $\cH_{AB}=\cH_A\ox\cH_B$, then $L_A=\tr_BL_{AB}$ is defined as its marginal on system $A$. Systems with the same
letter are assumed to be isomorphic: $A'\cong A$. We call a state a classical-quantum state if it is of the form
$\rho_{XA}=\sum_xp_X(x)\proj{x}_X\ox\rho_A^x$, where $p_X$ a probability distribution, $\{\ket{x}\}$ an orthonormal
basis of $\cH_X$, and $\rho^x_A\in\density{\cH_A}$. A linear map $\cN:\linear{\cH_A}\to\linear{\cH_B}$ maps operators
in system $A$ to operators in system $B$. $\cN_ {A\to B}$ is positive if $\cN_{A\to B}(\rho_A)\in\pos{\cH_B}$ whenever
$\rho_A\in\pos{\cH_A}$. Let $\id_A$ denote the identity map acting on system $A$. $\cN_{A\to B}$ is completely positive
if the map $\id_R\ox\cN_{A\to B}$ is positive for every reference system $R$. $\cN_{A\to B}$ is trace-preserving if
$\tr[\cN_{A\to B}(\rho_A)] = \tr\rho_A$ for all operator $\rho_A\in\linear{\cH_A}$. If $\cN_ {A\to B}$ is completely
positive and trace-preserving, we say that it is a quantum channel or quantum operation. A positive operator-valued
measure (POVM) is a set $\{\Lambda_m\}$ of operators satisfying $\forall m,\Lambda_m\geq0$ and $\sum_m\Lambda_m=\1$.

\paragraph*{Outline} The remainder of the paper is structured as follows. In Section~\ref{sec:informations}, we
introduce several different information of a quantum channel quantifying its ability in establishing correlation
assisted by the preshared entanglement. Properties and relations among these quantities are investigated and a
discord-type upper bound is derived. In Section~\ref{sec:semi-global}, we formally define the set of available encoding
operations and the classical communication task. Section~\ref{sec:capacity} is devoted to prove our main result -- a
capacity formula for the classical capacity defined in the last section. In Section~\ref{sec:examples} we consider
various quantum channels of interests and show that their classical capacities have analytically expression. We
conclude in Section~\ref{sec:conclusion} with some open problems.

\section{Information of a quantum channel}\label{sec:informations}

Let $\rho\in\subdensity{\cH}$ and $\sigma\in\pos{\cH}$ such that the support of $\rho$ is contained in the support of
$\sigma$. The quantum relative entropy is defined as. The quantum relative entropy is defined as
\begin{equation}
  \Rel\left(\rho\rel\sigma\right) := \tr\left[\rho(\log\rho - \log\sigma)\right],
\end{equation}
where logarithms are in base $2$ throughout this paper. The quantum entropy of $\rho$ is defined as
$\Shannon(\rho):=-\tr\rho\log\rho$. Let $\rho_{AB}\in\subdensity{\cH_A\ox\cH_B}$. 
The quantum mutual information and
conditional entropy of $\rho_{AB}$ are defined, respectively, as
\begin{IEEEeqnarray}{rCl}
  \Mutual\left(A{:}B\right)_\rho &:=& \Rel\left(\rho_{AB}\rel\rho_A\ox\rho_B\right), \\
  \Shannon\left(A{\vert}B\right)_\rho &:=& -\Rel\left(\rho_{AB}\rel\1_A\ox\rho_B\right).
\end{IEEEeqnarray}
The quantum mutual information of
$\rho_{AB}$ is defined as
\begin{equation}
  \Mutual\left(A{:}B\right)_\rho := \Rel\left(\rho_{AB}\rel\rho_A\ox\rho_B\right),
\end{equation}
while the conditional entropy of $\rho_{AB}$ is defined as
\begin{equation}
  \Shannon\left(A{\vert}B\right)_\rho := -\Rel\left(\rho_{AB}\rel\1_A\ox\rho_B\right).
\end{equation}

Let $\cN_{A\to B}$ be a quantum channel. The Holevo information of $\cN$ is defined as
\begin{equation}
  \chi(\cN) := \max_{\sigma_{XB}} \Mutual(X{:}B)_\sigma,
\end{equation}
where the maximization is taken over all classical-quantum states of the form
\begin{equation}
  \sigma_{XB} := \sum_xp_X(x)\proj{x}_X\ox\cN_{A\to B}(\rho_A^x),
\end{equation}
$p_X$ is \textit{a priori} probability distribution over alphabet $\cX$ and $\{\rho_A^x\}$ is a set of quantum states.
The mutual information of $\cN$ is defined as
\begin{equation}
  \Mutual(\cN) := \max_{\rho_A}\Mutual(\cN\vert\rho),    
\end{equation} 
where the maximization is taken over all quantum states in system $A$ and $\Mutual(\cN\vert\rho)$ is the mutual
information of $\cN$ w.r.t. (with respect to) the input state $\rho_A$:
\begin{equation}\label{eq:mutual-wrt-rho}
    \Mutual(\cN_{A\to B}\vert\rho_A) := \Mutual(A'{:}B)_\sigma,
\end{equation}
$\sigma_{A'B}=\cN_{A\to B}(\varphi_{A'A})$, and $\varphi_{A'A}$ is a purification of $\rho_A$.

The Holevo information and the mutual information represent two extremes of a quantum channel's ability to preserve
correlations: the former characterizes the ability to preserve the correlation without entanglement assistance, while
the latter characterizes the ability to preserve the correlation with unlimited entanglement assistance. Motivated by
this observation, we are interested in the ability of a quantum channel to preserve correlation assisted by limited
entanglement. We define two information measures aiming to quantify this ability.

\begin{definition}[Limited entanglement-assisted Holevo and mutual information of quantum channel]
\label{def:rho-mutual-information} 
Let $\rho_{E_AE_B}$ be a preshared bipartite state among Alice and Bob and let $\cN_{A\to B}$ be a quantum channel from
Alice to Bob. The $\rho_{E_AE_B}$-assisted Holevo information of $\cN$ is defined as
\begin{equation}\label{eq:rho-Holevo-information}
    \chi_{\rho_{E_AE_B}}\left(\cN\right) := \max_{\omega_{XBE_B}}\Mutual\left(X{:}BE_B\right)_\omega,
\end{equation}
while the $\rho_{E_AE_B}$-assisted mutual information of $\cN$ is defined as
\begin{equation}\label{eq:rho-mutual-information}
    \Mutual_{\rho_{E_AE_B}}\left(\cN\right) :=  \max_{\omega_{XBE_B}}\Mutual\left(XE_B{:}B\right)_\omega,
\end{equation}
where both maximizations are taken over classical-quantum states of the form
\begin{equation}\label{eq:omega_XBEB}
  \omega_{XBE_B} := \sum_x p_X(x) \proj{x}_X \ox \cN_{A\to B}\circ\cE^x_{E_A\to A}\left(\rho_{E_AE_B}\right),
\end{equation}
$p_X$ \textit{a priori} probability distribution over alphabet $\cX$ and $\{\cE^x_{E_A\to A}\}$ a set of encoding
channels.
\end{definition}
\begin{remark}
Note that $\chi_\rho(\cN)$ was previously defined and studied in~\cite{watrous2018theory}. However,
$\Mutual_\rho\left(\cN\right)$ is a new quantity to the best of our knowledge. This new definition is similar in the
form to $\chi_\rho(\cN)$ but using different partition with respect to which the mutual information is evaluated.
\end{remark}

We can use the quantum relative entropy ``distance'' between density operators to give the above defined four
information theoretic quantities a geometric and unified view. As we will see, these min-max formulas turn out to be
extremely helpful. The proof is given in Appendix~\ref{appx:prop:rel-characterization}.

\begin{proposition}\label{prop:rel-characterization}
It holds that
\begin{IEEEeqnarray}{rCl}
  \chi(\cN) &=& \min_{\sigma_B}\max_{\rho_A}\Rel\left(\cN_{A\to B}(\rho_A)\rel\sigma_B\right),\label{eq:hol-rel} \\
  \Mutual(\cN) &=& \min_{\sigma_B}\max_{\rho_A}
                  \Rel\left(\cN_{A\to B}(\varphi_{AA'})\rel\sigma_B\ox\varphi_{A'}\right),\label{eq:mut-rel} \\
  \chi_\rho(\cN) &=&  \min_{\sigma_{BE_B}}\max_{\cE_{E_A\to A}}
                    \Rel\left(\cN_{A\to B}\circ\cE_{E_A\to A}(\rho_{E_AE_B})\rel\sigma_{BE_B}\right),
                    \label{eq:chi-ea-rel} \\
  \Mutual_\rho(\cN) 
&=& \min_{\sigma_B}\max_{\cE_{E_A\to A}}
  \Rel\left(\cN_{A\to B}\circ\cE_{E_A\to A}(\rho_{E_AE_B})\rel\sigma_B\ox\rho_{E_B}\right),\label{eq:mutual-rel}
\end{IEEEeqnarray}
where $\varphi_{AA'}$ is a purification of $\rho_A$ and $\cE_{E_A\to A}$ ranges over all quantum channels
from $E_A$ to $A$.
\end{proposition}

The only difference between $\chi_\rho(\cN)$ and $\Mutual_\rho(\cN)$ lies in with which two parties we measure the
correlation -- for the former we measure the correlation w.r.t. system cut $X{:}BE_B$, while for the latter we measure
the correlation w.r.t. system cut $XE_B{:}B$. Interestingly enough, the new cut induces a larger correlation measure.
To show this, we need the following lemma, which is proved in Appendix~\ref{appx:lemma:difference}.
\begin{lemma}\label{lemma:difference}
For a state $\omega_{XBE_B}$ defined in~\eqref{eq:rho-mutual-information},
we have $\Mutual(XE_B{:}B)_\omega - \Mutual(X{:}BE_B)_\omega = \Mutual(B{:}E_B)_\omega$.
\end{lemma}

By taking the maximum w.r.t. $\omega_{XBE_B}$, Lemma~\ref{lemma:difference} yields the following relation.
\begin{proposition}\label{prop:chi-mutual-ea}
It holds that $\chi_\rho(\cN) \leq \Mutual_\rho(\cN)$. The equality holds if and only if the systems $E_B$ and $B$ are
independent under the optimal distribution $p_X$.
\end{proposition}

The BSST theorem~\cite{bennett1999entanglement,bennett2002entanglement} emphasized that $\Mutual(\cN)$ characterizes
the channel $\cN$'s ultimate ability to establish correlation, when unlimited entanglement is available. We want to
know if it is possible for $\Mutual_\rho(\cN)$ to reach $\Mutual(\cN)$ when $\rho$ is sufficiently entangled. We obtain
the following relation between $\Mutual_\rho(\cN)$ and $\Mutual(\cN)$. Especially, we give a necessary and sufficient
condition under which these two quantities are equal. The proof is deferred to
Appendix~\ref{appx:prop:mutual-relation}. Assume that $\varphi^\star_{A'A}$ is a state achieving $\Mutual(\cN)$
w.r.t.~\eqref{eq:mut-rel}.

\begin{proposition}\label{prop:mutual-relation}
It holds that $\Mutual_\rho(\cN)\leq\Mutual(\cN)$. The equality holds if and only if $\rho=\varphi^\star_{A'A}$.
\end{proposition}

Inspecting Propositions~\ref{prop:chi-mutual-ea} and~\ref{prop:mutual-relation}, we obtain a necessary and sufficient
condition under which $\chi_{\varphi^\star}(\cN)$ is equal to $\Mutual_{\varphi^\star}(\cN)$. The proof is given in
Appendix~\ref{appx:prop:chi-mutual-equal-condition}.
\begin{proposition}\label{prop:chi-mutual-equal-condition}
The equality of $\chi_{\varphi^\star}(\cN)\leq\Mutual_{\varphi^\star}(\cN)=\Mutual(\cN)$ holds if and only if the
reduced state $\tr_{A'}\varphi^\star_{A'A}$ is completely mixed on its support.
\end{proposition}

\subsection*{Upper bounds}

It is known that separable states are useless for classical
communication~\cite{bauml2019every,prevedel2011entanglement} while maximally entangled states double the classical
capacity of a noiseless channel using superdense coding~\cite{bennett1992communication}. These two extreme cases imply
that some states can improve a channel's classical communication ability while others cannot. This motivates the
question of to what extend a given bipartite quantum state $\rho_{E_AE_B}$ can enhance a channel's classical
communication capability, that is, how large the gap between $\Mutual_\rho(\cN)$ and $\chi(\cN)$ can be for an
arbitrary $\cN$?

It turns out that such an enhancement (if possible) is upper bounded by the \textit{discord of formation} of
$\rho_{E_AE_B}$, a quantity obtained from the relative entropy of quantum
discord~\cite{modi2010unified,modi2012classical} using the convex-roof construction~\cite{uhlmann1998entropy}. This
technique was previously applied to define the entanglement of
formation~\cite{bennett1996mixed,wootters1998entanglement} form the relative entropy of
entanglement~\cite{vedral1997quantifying}.
\begin{definition}[\cite{modi2010unified}]
Let $\rho_{AB}$ be a bipartite quantum state. The relative entropy of discord of $\rho_{AB}$ is defined as
\begin{equation}
  D_R(\rho_{AB}) := \min_{\{\proj{\phi_y}\}}\Rel\left(\rho_{AB}\rel\sum_yp_y\rho^y_A\ox\proj{\phi_y}_B\right),
\end{equation}
where the minimization is taken over all orthonormal bases $\{\proj{\phi_y}\}$ of system $B$,
$p_y:=\tr\bra{\phi_y}\rho_{AB}\ket{\phi_y}$, and $\rho^A_y:=\bra{\phi_y}\rho_{AB}\ket{\phi_y}/p_y$.
\end{definition}

Note that when $\varphi_{AB}$ is pure, $D_R(\varphi_{AB})$ evaluates to the entropy of the reduced state $\varphi_A$,
i.e., $D_R(\varphi_{AB}) = \Shannon(A)_\varphi$. Now we are ready to define a new discord measure -- the discord of
formation.
\begin{definition}[Discord of formation]
Let $\rho_{AB}$ be a bipartite quantum state. The discord of formation of $\rho_{AB}$ is defined as
\begin{equation}\label{eq:discord-of-formation}
  D_F(\rho_{AB}) := \min_{\rho_{AB} = \sum_xp_X(x)\rho_{AB}^x}\sum_xp_X(x)D_R\left(\rho_{AB}^x\right),
\end{equation}
where the minimization is taken over all possible probability distributions $p_X$ and choices of $\rho^x_{AB}$ such
that $\rho_{AB} = \sum_xp_X(x)\rho_{AB}^x$.
\end{definition}
If the minimization is restricted to pure state decompositions $\rho_{AB} = \sum_xp_X(x)\varphi_{AB}^x$
in~\eqref{eq:discord-of-formation}, we recover the definition of the entanglement of formation
$E_F$~\cite{bennett1996mixed}. As so, for arbitrary quantum states $\rho_{AB}$,
\begin{equation}\label{eq:D_F-E_F}
    D_F(\rho_{AB}) \leq E_F(\rho_{AB}).
\end{equation}

In Appendix~\ref{appx:proof-upper-bound} we show the following.
\begin{proposition}\label{prop:upper-bound}
Let $\rho_{E_AE_B}$ be a preshared bipartite state among Alice and Bob and let $\cN_{A\to B}$ be a quantum channel from
Alice to Bob. It holds that
\begin{equation}\label{eq:upper-bound}
\chi_\rho(\cN) - \chi(\cN) \leq \Mutual_\rho(\cN) - \chi(\cN) \leq D_F(\rho_{E_AE_B}).
\end{equation}
\end{proposition}

Using the fact that the entanglement of formation is faithful~\cite{bennett1996mixed}, we easily recover the fact that
separable states are useless for classical communication from Proposition~\ref{prop:upper-bound}. We remark that the
validity of the converse statement -- that every entangled state is helpful for classical communication -- is still
open~\cite{bauml2019every}.

\begin{corollary}\label{coro:mutual-min}
Let $\cN_{A\to B}$ be a quantum channel. It holds that
\begin{equation}
  \forall \rho_{E_AE_B}\in\SEP(E_A{:}E_B),\; \chi_\rho(\cN) = \Mutual_\rho(\cN) = \chi(\cN),
\end{equation}
where $\SEP(E_A{:}E_B)$ is the set of separable states of the composite system $E_AE_B$.
\end{corollary}

\section{Classical communication using semi-global operations}\label{sec:semi-global}

Throughout this section, we assume $\rho_{E_AE_B}$ a bipartite state preshared between Alice and Bob and $\cN_{A\to B}$
a quantum channel from Alice to Bob. We will describe the $\rho_{E_AE_B}$-assisted classical communication over
$\cN_{A\to B}$ such that only local operations and global permutations are allowed when encoding the message. To begin
with, we formally define the operations composed of local operations and global permutation. Such operations will be
termed as \textit{semi-global operations}.

\paragraph*{Semi-global operation} 

\begin{figure}[htbp]
  \centering
  \includegraphics[width=0.5\textwidth]{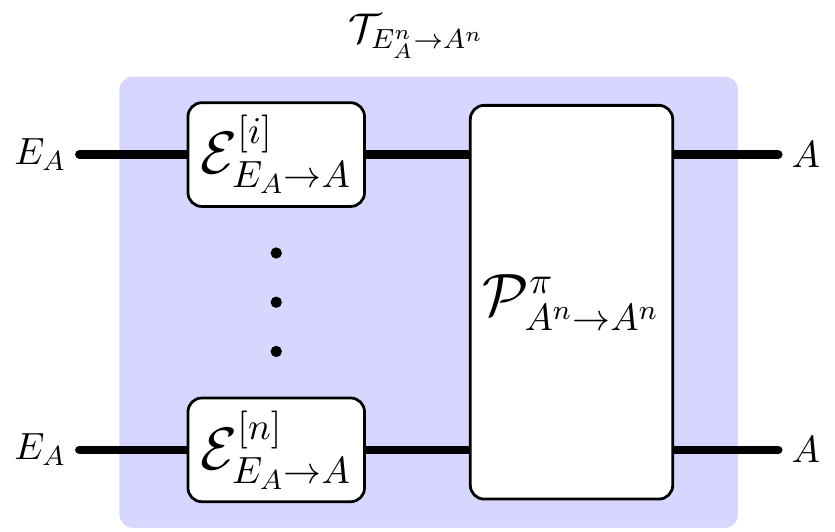}
  \caption{A semi-global operation $\cT_{E_A^{n} \to A^{n}}$. $\cT_{E_A^{n} \to A^{n}}$ is called semi-global if it can
      be decomposed as $n$ local channels $\cE^{[i]}_{E_A\to A}$ operating on $n$ different local systems and followed
      by a global permutation channel $\cP^\pi_{A^n\to A^n}$.}
\label{fig:semi-global-operation}
\end{figure}

Let $n$ be a positive integer. Let $S(n)$ be the set of permutations $\pi: [n]\to[n]$. Let $\pi\in S(n)$ be a
permutation and $\cP^\pi_{A^n\to A^n}$ be the permutation channel from $A^n$ to $A^n$ induced by $\pi$. Such a channel
reorders the output systems according to $\pi$. A semi-global operation from $E_A^n$ to $A^{n}$ is $n$ parallel local
channels from $E_A$ to $A$, followed by a permutation channel on $A^{n}$. Formally, a channel $\cT_{E_A^{n}\to A^{n}}$
is semi-global if there exists a set of local channels $\{\cE^{[i]}_{E_A\to A}\}_{i=1}^n$ and a permutation channel
$\cP^{\pi}_{A^n\to A^n}$ such that
\begin{equation}\label{eq:semi-global-channel}
\cT_{E_A^{n} \to A^{n}}(\cdot) :=  \cP^{\pi}_{A^n\to A^n}\circ
          \bigotimes_{i=1}^n\cE^{[i]}_{E_A\to A}(\cdot).
\end{equation}
Here the supscript $[i]$ indicates that the channel operates on the $i$-th input system. See
Fig.~\ref{fig:semi-global-operation} for illustration. At first glance, \eqref{eq:semi-global-channel} seemingly
does not cover all operations composed of local operations and global permutations. In Appendix~\ref{appx:semi-global}
we consolidate the compact definition of~\eqref{eq:semi-global-channel} does so. We denote by $\mathscr{T}_{E_A^n\to
A^n}$ the set of semi-global operations from $E_A^{n}$ to $A^{n}$. When $n=1$, $\mathscr{T}_{E_A\to A}$ reduces to the
set of quantum channels from $E_A$ to $A$.

\paragraph*{$\rho$-assisted classical communication using semi-global operations}

\begin{figure}[htbp]
  \centering
  \includegraphics[width=0.8\textwidth]{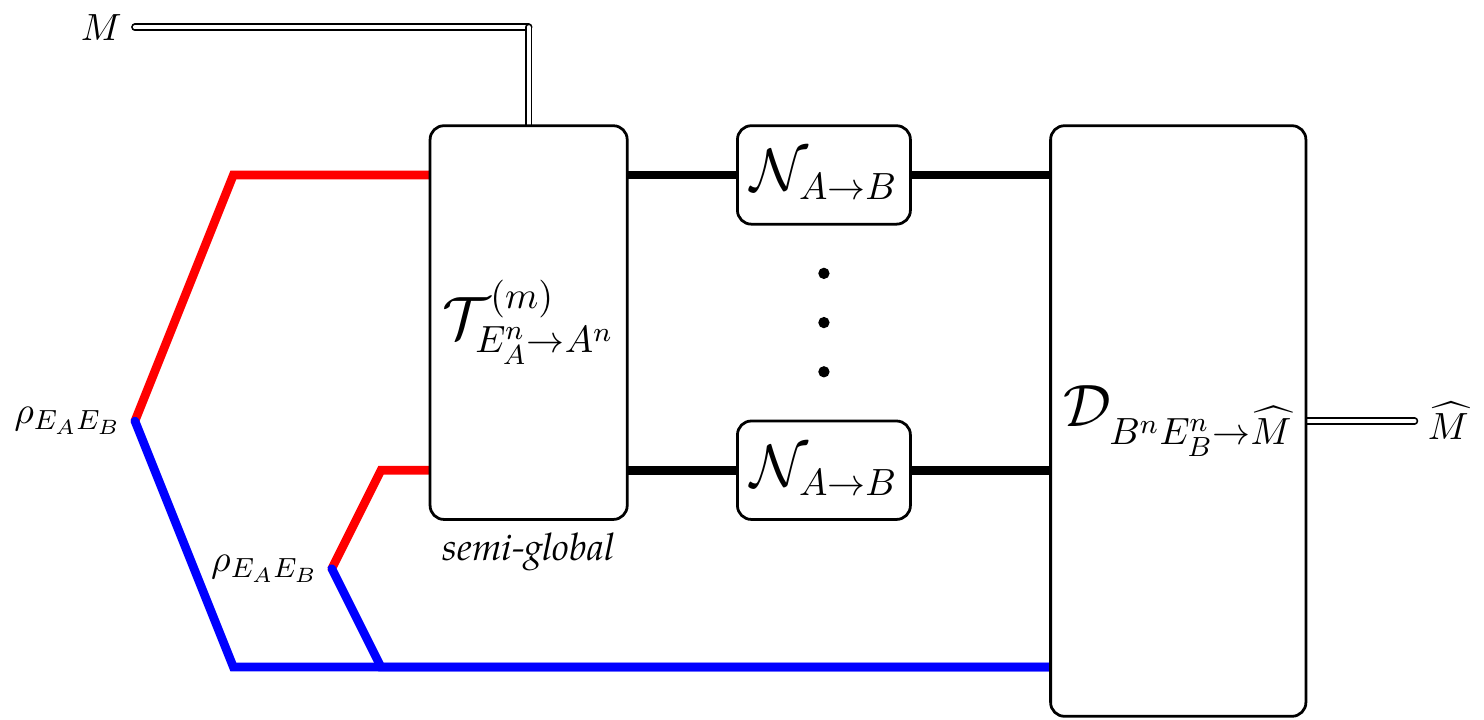}
  \caption{Semiproduct channel coding framework of blocklength $n$. For each message $m\in\cM_n$, Alice encodes $m$
    into her part of the state $\rho_{E_AE_B}^{\ox n}$ by performing a semi-global encoding operation $\cT^m_{E_A^n\to
    A^n}$. After receiving the state through $n$ independent uses of $\cN_{A\to B}$, Bob performs a decoding operation
    $\cD_{B^nE_B^n\to\wh{M}}$ to infer the encoded message $m$.}
\label{fig:semi-global-coding-framework}
\end{figure}

The task is to transmit classical message as much as possible from Alice to Bob, using multiple copies of $\rho$
through many independent uses of $\cN$, under the constraint that a copy of $\rho$ is consumed per channel use and
$\rho$ cannot be distributed among more than one channel. Consider now a channel coding of blocklength $n$. Alice
selects some message $m$ from the alphabet $\cM_n$, whose size is $M_n$. Let $M$ denote the random variable
corresponding to Alice's choice. She applies a semi-global operation $\cT^m_{E_A^n\to A_n}$ to her share of the state
$\rho^{\ox n}_{E_AE_B}$ depending on message $m$. In this way, she encodes $m$ into the preshared quantum states. This
is called \textit{semi-global coding} since only semi-global operations are allowed on Alice's side. After encoding,
Alice and Bob share the state
\begin{equation}
    \sigma^m_{A^nE_B^n} 
:=  \cT^m_{E_A^n\to A^n}\left(\rho^{\ox n}_{E_AE_B}\right)
 =  \cP^m_{A^n\to A^n}\circ\bigotimes_{i=1}^n\cE^{[i]\vert m}_{E_A\to A}\left(\rho_{E_AE_B}\right),
\end{equation}
where $\{\cE^{[i]\vert m}_{E_A\to A}\}$ and $\cP^m$ are chosen such that $\cT^m_{E_A^n\to A^n}$ can be decomposed
as~\eqref{eq:semi-global-channel}. Note that $\cT^m$ is message $m$ dependent and so is $\cE^{[i]\vert m}$ and $\cP^m$.
After encoding, Alice sends her encoded state to Bob, through $n$ independent uses of  $\cN_{A\to B}$, leading to the
state
\begin{equation}
    \omega^m_{B^nE_B^n} 
:= \cN_{A\to B}^{\ox n}\left(\sigma^m_{A^nE_B^n}\right)
= \cN^{\ox n}\circ\cP^m\circ\bigotimes_{i=1}^n\cE^{[i]\vert m}\left(\rho\right).
\end{equation}
On receiving the state, Bob performs a measurement $\cD:=\{\Lambda_{\wh{m}}\}_{\wh{m}\in\cM_n}$ on
$\omega^m_{B^nE_B^n}$ to infer the encoded message $m$. Fig.~\ref{fig:semi-global-coding-framework} depicts this
semi-global coding protocol for the $\rho$-assisted classical communication. The protocol $(n,\cT,\cD)$ is called a
\textit{semi-global coding protocol of blocklength $n$} for the state-channel pair $(\rho,\cN)$.

Let $\wh{M}$ be the random variable corresponding to the output of Bob's decoding, representing Bob's inferred message.
The decoding operation leads to the classical state
\begin{equation}
\gamma_{M\wh{M}}  := \frac{1}{M_n}\sum_{m,\wh{m}}p_{\wh{m}\vert m}\proj{m}_{M}\ox\proj{\wh{m}}_{\wh{M}},
\end{equation}
where the conditional decoding probability obeys
\begin{equation}
  p_{\wh{m}\vert m} := \tr\left[\Lambda_{\wh{m}}\omega^m_{B^nE_B^n}\right]. 
\end{equation}
A decoding error occurs if the output $\wh{m}$ is not equal to the input $m$. The probability that Bob successfully
decodes $m$ is given by
\begin{equation}\label{eq:bob-success}
    \Pr\left\{\wh{M} = m \middle\vert M= m\right\} = p_{m \vert m}.
\end{equation}
As a result, the probability of error for a particular message $m$ is
\begin{equation}
    p_e(m) := 1 - p_{m \vert m} = \tr\left[(\1 - \Lambda_{m})\omega^m_{B^nE_B^n}\right].
\end{equation}

We need to quantify the performance of the protocol $(n,\cT,\cD)$. One commonly adopted way to quantify the performance
of the protocol is to compute the average probability of error that the decoded message $\wh{M}$ is not equal to the
encoded message $M$:
\begin{equation}
    e(n,\cT,\cD) := \frac{1}{M_n}\sum_{m\in\cM_n} p_e(m).
\end{equation}
In general, smaller error probability indicates better protocol. However, in order to make the error probability small,
one can only encode classical message with a smaller size. This motivates us to define another quantity that
quantitatively measure the size of the encoded message. We define the coding rate of the protocol as
\begin{equation}
    r\left(n,\cT,\cD\right) := \frac{1}{n}\log M_n.
\end{equation}
It measures how many bits of classical message can be transmitted per state and channel use.

Let $\varepsilon\in[0,1)$ fixed. A coding protocol $(n,\cT,\cD)$ is said to be an $(n,R,\varepsilon)$-code for
$(\rho,\cN)$, if the protocol satisfies the following two conditions:
\begin{itemize}
   \item Coding rate condition: $R = r(n,\cT,\cD)$; and
   \item Performance condition: $e(n,\cT,\cD) \leq \varepsilon$.
\end{itemize}
Intuitively, these two conditions state that the coding protocol $(n,\cT,\cD)$ can transmit classical message at rate
$R$ with probability of error at most $\varepsilon$. Let $R\in\bRp$ fixed. If for \textit{arbitrary} $\delta>0$, there
always exists an $(n,R-\delta,\varepsilon)$-code for $(\rho,\cN)$ when $n$ is sufficiently large, we say this rate $R$
is $\varepsilon$-\textit{achievable}. The $\rho$-assisted $\varepsilon$-classical capacity of $\cN$ is defined to be
the supremum of all achievable rates.

\begin{definition}[$\rho$-assisted classical capacity with semi-global operations]
Let $\varepsilon\in[0,1)$. The $\rho$-assisted $\varepsilon$-classical capacity of $\cN$, when semi-global operations
is available, is defined as
\begin{equation}
  C^{\pi,\varepsilon}_{\rho}\left(\cN\right) 
:= \sup\left\{R: \textit{rate $R$ is $\varepsilon$-achievable for $(\rho,\cN)$ using semi-global operations}\right\}.
\end{equation}
\end{definition}
\noindent The supscript $\pi$ in $C^{\pi,\varepsilon}_\rho$ refers to permutation and indicates that the capacity is
defined by using only semi-global operations, and the supscript $\varepsilon$ indicates that the decoding error
probability is upper bounded by constant $\varepsilon$. By definition, it is easy to see $C^{\pi,\varepsilon}_{\rho}$
is monotonic in $\varepsilon$ in the sense that
\begin{equation}\label{eq:C-monotonicity}
  \varepsilon \leq \varepsilon' \quad\Rightarrow\quad 
  C^{\pi,\varepsilon}_{\rho}\left(\cN\right) \leq C^{\pi,\varepsilon'}_{\rho}\left(\cN\right).
\end{equation}

\paragraph*{Holevo information using semi-global operations}

Let $n\in\bNp$. We define the $n$-th $\rho$-assisted Holevo information of $\cN$,
using only semi-global operations, as
\begin{equation}
  \chi^{\pi,n}_{\rho}\left(\cN\right)
:= \frac{1}{n}\max_{\omega_{XB^nE_B^n}}\Mutual\left(X{:}B^nE_B^n\right)_\omega,
\end{equation}
where
\begin{equation}\label{eq:omega-XBnEBn}
  \omega_{XB^nE_B^n} := 
  \sum_{x\in\cX} p_x \proj{x}_X \ox \cN^{\ox n}_{A\to B}\circ\cT^x_{E_A^n\to A^n}\left(\rho^{\ox n}_{E_AE_B}\right),
\end{equation}
$\{p_x\}$ is \textit{a priori} probability distribution over the message space $\cX$, and
$\{\cT^x\in\mathscr{T}_{E_A^n\to A^n}\}$ is a set of semi-global operations.
Correspondingly, the regularized $\rho$-assisted Holevo information of a quantum channel,
using only semi-global operations, is defined as
\begin{equation}
  \ol{\chi}^{\pi}_{\rho}\left(\cN\right) := \limsup_{n\to\infty}\chi^{\pi,n}_{\rho}\left(\cN\right).
\end{equation}
The regularized $\rho$-assisted Holevo information is a lower bound on $C^{\pi,\varepsilon}_\rho$.
\begin{proposition}\label{prop:capacity-regularization}
Let $\varepsilon\in[0,1)$.
It holds that $C^{\pi,\varepsilon}_\rho\left(\cN\right)\geq\ol{\chi}^{\pi}_{\rho}\left(\cN\right)$.
\end{proposition}
\begin{IEEEproof}
By the achievability part of the Holevo-Schumacher-Westmoreland theorem~\cite{schumacher1997sending,holevo1998capacity}
(see also~\cite[Chapter 4]{hayashi2016quantum} for a thorough discussion), it holds that the regularized
$\rho$-assisted Holevo information $\ol{\chi}^{\pi}_{\rho}\left(\cN\right)$ is an achievable rate with asymptotically
vanishing error probability, that is
\begin{equation}
    C^{\pi,0}_\rho\left(\cN\right)\geq\ol{\chi}^{\pi}_{\rho}\left(\cN\right).
\end{equation}
On the other hand,
the monotonicity~\eqref{eq:C-monotonicity} guarantees the above inequality holds for arbitrary $\varepsilon\in[0,1)$.
\end{IEEEproof}

\paragraph*{Classical communication using product operations}

Assume now that in the above channel coding framework, the permutation over systems $A^{n}$ is not allowed. In this
case we can also define a classical capacity using only product operations.
\begin{definition}[$\rho$-assisted classical capacity with product operations]
The $\rho$-assisted $\varepsilon$-classical capacity of $\cN$, when only product operations is available, is defined as
\begin{equation}
  C^{\ox,\varepsilon}_\rho\left(\cN\right) 
:= \sup\left\{R: \textit{rate $R$ is $\varepsilon$-achievable for $(\rho,\cN)$ using product operations}\right\}.
\end{equation}
\end{definition}
The supscript $\ox$ in $C^{\ox,\varepsilon}_\rho$ indicates that the capacity is defined using only product operations. 

By the achievability part of the Holevo-Schumacher-Westmoreland
theorem~\cite{schumacher1997sending,holevo1998capacity}, $\chi_\rho(\cN)$ is an achievable rate for
$C^{\ox,\varepsilon}_\rho\left(\cN\right)$. On the other hand, since the set of quantum channels $\{\cE_{E_A\to A}\}$
is compact, it then follows that $\chi_\rho(\cN)$ is a strong converse bound for
$C^{\ox,\varepsilon}_\rho\left(\cN\right)$~\cite[Chapter 4]{hayashi2016quantum}. To summarize, we obtain the following.

\begin{proposition}\label{prop:capacity-product}
Let $\varepsilon\in[0,1)$. It holds that $C^{\ox,\varepsilon}_\rho\left(\cN\right) = \chi_\rho(\cN)$.
\end{proposition}

\section{The capacity formula}\label{sec:capacity}

In this section we will show that when pure state $\varphi_{E_AE_B}$ is available, we can derive a ``single-letter''
formula for the capacity $C^{\pi,\varepsilon}_{\varphi}\left(\cN\right)$ --- it is given exactly by
$\Mutual_\varphi(\cN)$, the $\varphi$-assisted mutual information of channel $\cN$. This result - together with
Propositions~\ref{prop:chi-mutual-ea} and~\ref{prop:capacity-product} - reveals the power of global permutation in
classical communication: it can increase the communication rate compared to the case when only product operations is
available. 

We first consider the converse part. We show that $\Mutual_\rho(\cN)$ is a strong converse bound for arbitrary
$\rho_{E_AE_B}$-assisted classical communication using semi-global operations.
\begin{lemma}[Strong converse]\label{lemma:converse}
The inequality $C^{\pi,\varepsilon}_\rho\left(\cN\right)\leq\Mutual_\rho(\cN)$
holds for $\varepsilon\in[0,1)$. 
\end{lemma}
Combining Proposition~\ref{prop:capacity-product} and Lemma~\ref{lemma:converse}, 
we have for arbitrary $\varepsilon\in[0,1)$ that
\begin{equation}
\chi_\rho(\cN) = C^{\ox,\varepsilon}_\rho\left(\cN\right)
\leq C^{\pi,\varepsilon}_\rho\left(\cN\right)\leq \Mutual_\rho(\cN).
\end{equation}
Due to Proposition~\ref{prop:chi-mutual-ea}, the equalities in both inequalities hold when there exists a distribution
$p_X$ on the set $\{\cE_{E_A\to A}^x\}$ for which $E_B$ and $B$ independent and $I(X E_B{:}B)_\omega=
\Mutual_\rho(\cN)$. Under this equality condition, local product operations are as powerful as the semi-global
operations in $\rho$-assisted classical communication.

\begin{lemma}[Achievability]\label{lemma:achievability}
The inequality $C^{\pi,\varepsilon}_\varphi\left(\cN\right)\geq\Mutual_\varphi(\cN)$
holds for $\varepsilon\in[0,1)$ when $\varphi$ is pure.
\end{lemma}

As a corollary of the above two Lemmas~\ref{lemma:converse} and~\ref{lemma:achievability}, we obtain the following
capacity formula.

\begin{theorem}\label{thm:capacity}
The equation $C^{\pi,\varepsilon}_\varphi\left(\cN\right) = \Mutual_\varphi(\cN)$ holds for 
$\varepsilon\in[0,1)$ when $\varphi$ is pure.
\end{theorem}
Combining Proposition~\ref{prop:capacity-product} and Theorem~\ref{thm:capacity},
we have for arbitrary $\varepsilon\in[0,1)$ that
\begin{equation}
    \chi_\varphi(\cN) = C^{\ox,\varepsilon}_{\varphi}(\cN) 
\leq C^{\pi,\varepsilon}_\varphi\left(\cN\right) = \Mutual_\varphi(\cN).
\end{equation}
Therefore, whenever the equality condition given in Proposition~\ref{prop:chi-mutual-ea} does not hold, global
permutations increase the communication rate compared to the case when only product operations is available.

\subsection{Achievability}\label{sec:achievability}

This section aims to prove Lemma~\ref{lemma:achievability}. More specifically, we will construct a sequence of state
ensembles induced by semi-global operations for which the regularized Holevo information satisfies
\begin{equation}\label{eq:I-varphi-achievability}
    \ol{\chi}^{\pi}_{\varphi}\left(\cN\right) \geq \Mutual_\varphi(\cN).
\end{equation}
This fact together with Proposition~\ref{prop:capacity-regularization} implies Lemma~\ref{lemma:achievability}.

Let $\cX\equiv\{a_1,\cdots, a_{\vert\cX\vert}\}$ be an alphabet of size $\vert\cX\vert$. Assume $\{p_X(x),
\cE^x_{E_A\to A}\}_ {x\in\cX}$ achieves $\Mutual_\varphi(\cN)$. Define the following quantum states:
\begin{IEEEeqnarray}{rCl}
  \sigma^x_B      &:=& \cN_{A\to B}\circ\cE^x_{E_A\to A}(\varphi_{E_A}), \\
  \sigma^x_{BE_B} &:=& \cN_{A\to B}\circ\cE^x_{E_A\to A}(\varphi_{E_AE_B}), \\
  \sigma_{XBE_B}  &:=& \sum_xp_x\proj{x}_X\ox\cN_{A\to B}\circ\cE^x_{E_A\to A}(\varphi_{E_AE_B})
                    = \sum_xp_x\proj{x}_X\ox\sigma^x_{BE_B}.
\end{IEEEeqnarray}
By assumption
\begin{equation}
  \Mutual_\varphi(\cN) = \Mutual(XE_B{:}B)_\sigma
                    = \Shannon(B)_\sigma + \Shannon(E_B)_\sigma - \Shannon(BE_B\vert X)_\sigma.
\end{equation}
The $n$-th tensor of $\sigma_{XBE_B}$ has the form
\begin{equation}
  \sigma_{X^nB^nE^n_B} = \sum_{x^n\in\cX^n}p_{X^n}(x^n)\proj{x^n}_{X^n}\ox\sigma_{B^nE^n_B}^{x^n},
\end{equation}
where $x^n \equiv x_1\cdots x_n$, 
\begin{equation}
p_{X^n}(x^n) \equiv \prod_{i=1}^np_X(x_i),\quad
\ket{x^n} \equiv \ket{x_1\cdots x_n},\quad
\sigma_{B^nE^n_B}^{x^n} \equiv \bigotimes_{i=1}^n\sigma_{BE_B}^{x_i}.
\end{equation}

Let $\rho_A$ be a quantum state with spectral decomposition $\rho_A=\sum_{z}\lambda_z\proj{z}_{A}$.
Let $\{Z^u_\rho\}$ be the set of Weyl operators in system $A$ w.r.t. to basis $\{\ket{z}\}$, i.e.,
\begin{equation}\label{eq:Z_u_rho}
    Z^u_\rho := \sum_{z=0}^{d_A-1}e^{2uz\pi i/d_A}\proj{z}_A,\; u = 0,\cdots, d - 1,
\end{equation}
where $d_A$ is the dimension of system $A$. We denote by $\cZ_\rho^u(\cdot):=
Z_\rho^u(\cdot)(Z_\rho^u)^\dagger$ the corresponding unitary channel. One can check that $\cZ^u_\rho(\rho)
= \rho$. 

Assume the reduced state $\varphi_{E_A}$ of $\varphi_{E_AE_B}$ has the spectral decomposition
$\varphi_{E_A}=\sum_{z}\lambda_z\proj{z}$. Let $Z$ be a random variable such that $p_Z(z)=
\bra{z}\varphi_{E_A}\ket{z} = \lambda_z$. Let $\varphi_{E_B\vert z}
:=\bra{z}\varphi_{E_AE_B}\ket{z}/\lambda_z$. Then
\begin{equation}\label{eq:Z-expectation}
  \frac{1}{d}\sum_u\cZ^u_{\varphi_{E_A}}(\varphi_{E_AE_B})
= \frac{1}{d}\sum_u(Z_\varphi^u\ox\1_{E_B}) \varphi_{E_AE_B} (Z_\varphi^u\ox\1_{E_B})
= \sum_zp_Z(z)\proj{z}_Z\ox\varphi_{E_B\vert z} \equiv \rho_{ZE_B}.
\end{equation}
That is, random phase changing operations $\cZ^u$ erase the entanglement in $\varphi_{E_AE_B}$, resulting a
classical-quantum state $\rho_{ZE_B}$, with both systems $E_A$ and $E_B$ dephased in their eigenbases.

Let $\Lambda_n$ be a random variable with alphabet $S(n)$ and probability distribution $p(\Lambda_n=\pi_n)=1/(n!)$.
$\Lambda_n$ represents the event of choosing a permutation randomly and uniformly from $S(n)$. Our achievability proof
makes use of the following lemma.
\begin{lemma}\label{lemma:permutation-existence}
Let $n\in\bNp$. Let $\cM_{E_A\to B}$ be an arbitrary quantum channel. Define the following states:
\begin{IEEEeqnarray}{rCl}
  \sigma_{BE_B} &:=& \cM_{E_A\to B}(\varphi_{E_AE_B}), \\
  \omega_{\Lambda_nU^nB^nE^n_B} 
&:=& \frac{1}{n!d^n}\sum_{u^n,\pi_n}\proj{\pi_n}_{\Lambda_n}\ox\proj{u^n}_{U^n}\ox 
        \cM^{\ox n}_{E_A\to B}\circ\cP^{\pi_n}\circ\cZ_\varphi^{u^n}(\varphi^{\ox n}_{E_AE_B}),\label{eq:lemma-state}
\end{IEEEeqnarray}
where $d$ is the dimension of system $E_A$. It holds that
\begin{equation}
  n\Mutual(B{:}E_B)_\sigma - \Mutual(\Pi_nU^n{:}B^nE_B^n)_{\omega} \leq d\log(n+1).
\end{equation}
\end{lemma}

\begin{IEEEproof}
For each $\pi_n$ and $u^n$, define the conditional state:
\begin{equation}
    \omega^{\pi_n,u^n}_{B^nE^n_B} := \cM^{\ox n}\circ\cP^{\pi_n}\circ\cZ_\varphi^{u^n}(\varphi^{\ox n}_{E_AE_B}).
\end{equation}
Then
\begin{equation}
  \omega_{\Lambda_nU^nB^nE^n_B} 
= \frac{1}{n!d^n}\sum_{u^n,\pi_n}\proj{\pi_n}_{\Lambda_n}\ox\proj{u^n}_{U^n}\ox\omega^{\pi_n,u^n}_{B^nE^n_B}.
\end{equation}
We have the following reduced states:
\begin{IEEEeqnarray}{rCl}
  \omega_{B^nE^n_B} 
&=&  \frac{1}{n!d^n}\sum_{\pi_n,u^n}\cM^{\ox n}\circ\cP^{\pi_n}\circ\cZ_\varphi^{u^n}(\varphi^{\ox n}_{E_AE_B}), \\
      \omega_{B^n} 
&=&  \frac{1}{n!d^n}\sum_{\pi_n,u^n}\cM^{\ox n}\circ\cP^{\pi_n}\circ\cZ_\varphi^{u^n}(\varphi^{\ox n}_{E_A})
      = \cM^{\ox n}(\varphi^{\ox n}_{E_A}) = \sigma_B^{\ox n},\label{eq:reduced-state-B} \\
      \omega_{E_B^n} 
&=& \sigma_{E_B}^{\ox n} = \varphi_{E_B}^{\ox n}.\label{eq:reduced-state-EB}
\end{IEEEeqnarray}
Since $\varphi_{E_AE_B}$ is pure and $Z^u_\varphi$ commutes with $\varphi_{E_A}$, we
can identify another unitary $\wt{Z}_\varphi^u$ on $E_B$ such that
\begin{equation}
    (Z^u_\varphi\ox\1_{E_B})\ket{\varphi_{E_AE_B}} = (\1_{E_A}\ox\wt{Z}^u_\varphi)\ket{\varphi_{E_AE_B}}.
\end{equation}
Also, for each permutation $\cP^{\pi_n}$ on $E_A^n$, 
we can always choose a permutation $\cP^{\pi'_n}$ on $E_B^n$ such that
\begin{equation}
  \left(\cP^{\pi_n}\ox\id_{E_B^n}\right)\left(\varphi_{E_AE_B}^{\ox n}\right)
= \left(\id_{E_A^n}\ox\cP^{\pi'_n}\right)\left(\varphi_{E_AE_B}^{\ox n}\right).
\end{equation}
The above two observations tell us that the operations $Z^u_\varphi$ and $\cP^{\pi_n}$ on system $E_A$ can
be exchanged to corresponding operations on system $E_B$ without altering the output state. As so
\begin{IEEEeqnarray}{rCl}
\Shannon\left(B^nE_B^n{\vert}\Lambda_nU^n\right)_{\omega}
&=&  \frac{1}{n!d^n}\sum_{\pi_n,u^n}\Shannon\left(B^nE_B^n\right)_{\omega^{\pi_n,u^n}} \\
&=&  \frac{1}{n!d^n}\sum_{\pi_n,u^n}
      \Shannon\left(\cM^{\ox n}_{E_A\to B}\circ\cP^{\pi_n}\circ\cZ^{u^n}(\varphi^{\ox n}_{E_AE_B})\right) \\
&=&  \frac{1}{n!d^n}\sum_{\pi'_n,u^n}
      \Shannon\left(\left(\id_{E^n_A}\ox\cP^{\pi'_n}\right)\circ
      \left(\id_{E^n_A}\ox\cZ^{u^n}\right)\circ\cM^{\ox n}_{E_A\to B}(\varphi^{\ox n}_{E_AE_B})\right) \\
&=&  \frac{1}{n!d^n}\sum_{\pi'_n,u^n}
      \Shannon\left(\cM^{\ox n}_{E_A\to B}(\varphi^{\ox n}_{E_AE_B})\right) \\
&=&  \Shannon\left(B^nE_B^n\right)_\sigma.\label{eq:ensemble-entropy}
\end{IEEEeqnarray}
Consider the following chain of inequalities:
\begin{IEEEeqnarray}{rCl}
&~&    \Mutual(B^n{:}E_B^n)_\sigma - \Mutual(\Lambda_nU^n{:}B^nE_B^n)_{\omega} \\
&=&  \Shannon(B^n)_\sigma + \Shannon(E_B^n)_\sigma - \Shannon(B^nE_B^n)_\sigma
    - \Shannon(B^nE_B^n)_\omega - \Shannon\left(B^nE_B^n{\vert}\Pi_nU^n\right)_{\omega} \\
&=&  \Shannon(B^n)_\omega + \Shannon(E_B^n)_\omega - \Shannon(B^nE_B^n)_\omega \\
&=&  \Mutual(B^n{:}E_B^n)_\omega \\
&\leq& \Mutual(E_A^n{:}E_B^n)_\tau,
\end{IEEEeqnarray}
where the last inequality follows from the data-processing inequality.
Let's go depth on the state $\tau_{E_A^nE_B^n}$:
\begin{IEEEeqnarray}{rCl}
      \tau_{E_A^nE_B^n}
&:=& \frac{1}{n!d^n}\sum_{\pi_n,u^n}\cP^{\pi_n}\circ\cZ_\varphi^{u^n}(\varphi^{\ox n}_{E_AE_B}) \\
&=&  \frac{1}{n!}\sum_{\pi_n}\cP^{\pi_n}\left(
      \frac{1}{d^n}\sum_{u^n}\cZ_\varphi^{u^n}(\varphi^{\ox n}_{E_AE_B})\right) \\
&=&  \frac{1}{n!}\sum_{\pi_n}\cP^{\pi_n}\left(\rho_{ZE_B}^{\ox n}\right),
\end{IEEEeqnarray}
where the last equality follows from~\eqref{eq:Z-expectation}. Let $\cT$ be the set of types for $Z^n$, $t\in\cT$
be a type, $T^{Z^n}_{t}$ be the type class corresponding to $t$, $d_t$ be the size of $T_t^{Z^n}$, and $p_t$ be the
probability of sequences in $T_t^{Z^n}$. We refer to~\cite{hayashi2016quantum,wilde2016quantum} for more information on
the concept of type and its applications. We have
\begin{IEEEeqnarray}{rCl}
\frac{1}{n!}\sum_{\pi_n}\cP^{\pi_n}\left(\rho_{ZE_B}^{\ox n}\right)
&=&  \frac{1}{n!}\sum_{\pi_n}\cP^{\pi_n}
      \left(\sum_{t\in \cT}\sum_{z^n\in T_t^{Z^n}}p_t\proj{z^n}_{Z^n}\ox\rho_{z^n}\right) \\
&=&  \sum_{t\in \cT}p_t\sum_{z^n\in T_t^{Z^n}}\proj{z^n}_{Z^n}\ox
      \frac{1}{n!}\sum_{\pi_n}\cP^{\pi_n}\left(\rho_{z^n}\right) \\
&=&  \sum_{t\in \cT}p_t\sum_{z^n\in T_t^{Z^n}}\proj{z^n}_{Z^n}\ox\rho_t,
\end{IEEEeqnarray}
where
\begin{equation}
    \rho_t := \frac{1}{d_t}\sum_{z^n\in T_t^{Z^n}}\rho_{z^n}.
\end{equation}
Set $q_t \equiv p_td_t$. Then $\sum_{t\in\cT}q_t=1$ and
\begin{IEEEeqnarray}{rCl}
  \Mutual\left(E_A^n{:}E_B^n\right)_{\tau}
&=& \Shannon\left(E_B^n\right)_{\tau} - \Shannon\left(E_B^n{\vert}E_A^n\right)_{\tau} \\
&=& \Shannon\left(\sum_tq_t\rho_t\right) - \sum_tq_t\Shannon(\rho_t) \\
&\leq& \Shannon(\{q_t\}) \label{eq:lemma8-tmp1} \\
&\leq& \log\vert\cT\vert \label{eq:lemma8-tmp2} \\
&\leq& d\log(n+1), \label{eq:lemma8-tmp3}
\end{IEEEeqnarray}
where~\eqref{eq:lemma8-tmp1} follows from the flip side of concavity of the
entropy~\cite[(11.79)]{nielsen2011quantum},~\eqref{eq:lemma8-tmp2} is the entropy dimension bound,
and~\eqref{eq:lemma8-tmp3} follows from an upper bound on the number of types~\cite[Property 14.7.1]{wilde2016quantum}.
We are done.
\end{IEEEproof}

Now we are ready to show the achievability part.

\begin{IEEEproof}[Proof of~\eqref{eq:I-varphi-achievability}] 
Our goal is to construct a set of signal states whose Holevo information is no less than $\Mutual_\varphi(\cN)$ when
$n$ is sufficiently large. Let $\cT$ be the set of types for $X^n$, $t\in\cT$ be a type, $T^{X^n}_{t}$ be the type
class corresponding to $t$, $d_t$ be the size of $T_t^{X^n}$, and $p_t$ be the probability of sequences in $T_t^{X^n}$.
Fix $t$. For each $a\in\cX$, define the following quantities
\begin{equation}\label{eq:permutation-size}
  \Lambda_{nt(a)} := (nt(a))!,\; \Lambda_a := \max_{t\in\cT}\Lambda_{nt(a)},\quad
  \bm{\Lambda}:=\prod_{a\in\cX}\Lambda_a.
\end{equation}
Let $\bm{\pi}:=\{\pi_1,\cdots,\pi_{\vert\cX\vert}\}$ be an instance of $\bm{\Lambda}$ such that each $\pi_i$ is an
instance of $\Lambda_a$. Define the conditional probability distribution $p_{\bm{\Lambda}\vert X^n}$ as
\begin{equation}\label{eq:conditional-probability}
\setlength{\nulldelimiterspace}{0pt}
p_{\bm{\Lambda}\vert X^n}(\bm{\pi}\vert x^n) =
\left\{\begin{IEEEeqnarraybox}[\relax][c]{l's}
\frac{1}{\Pi_{a\in\cX} \Lambda_{nt_{x^n}(a)}},& $\forall a\in\cX, \pi_a\leq \Lambda_{nt_{x^n}(a)}$\\
0,& otherwise%
\end{IEEEeqnarraybox}\right.
\end{equation}
This is an valid conditional probability distribution since $p_{\bm{\Lambda}\vert X^n}(\bm{\pi}\vert x^n)\geq 0$ and
\begin{equation}
  \forall x^n\in\cX^n,\; \sum_{\bm{\pi}} p_{\bm{\Lambda}\vert X^n}(\bm{\pi}\vert x^n) = 1.
\end{equation}

Fix $x^n$. Let $\wt{n}_a\equiv nt_{x^n}(a)$. We classify the $n$ input systems $E_A$ into $\vert\cX\vert$ groups based
on $x^n$ such that each group contains systems $E_A$ belonging to the same alphabet, that is,
\begin{equation}\label{eq:classification}
  E_A^n \mapsto \bigotimes_{a\in\cX} E_A^{\wt{n}_a}.
\end{equation}
For each conditional sequence $\bm{\pi}\vert x^n$, we define the following permutation operation:
\begin{equation}\label{eq:permutation-construction}
    \cP^{\bm{\pi}\vert x^n} := \bigotimes_{a\in\cX}\cP^{\pi_a}_{E_A^{\wt{n}_a}\to E_A^{\wt{n}_a}},
\end{equation}
where each permutation $\cP^{\pi_a}$ operates only on the group $a$ as classified in~\eqref{eq:classification} and
is indexed by $\pi_a$. The size of $S(\wt{n}_a)$ is $\Lambda_{\wt{n}_a}$, which is
smaller than $\Lambda_a$ by definition~\eqref{eq:permutation-size}. As so, $\pi_a$ is possibly out of range when
indexing the set of permutation channels from $E_A^{\wt{n}_a}$ to $E_A^{\wt{n}_a}$. To get rid of this problem,
we make the following convention:
\begin{equation}
\setlength{\nulldelimiterspace}{0pt}
\cP^{\pi_a}_{E_A^{\wt{n}_a}\to E_A^{\wt{n}_a}} =
\left\{\begin{IEEEeqnarraybox}[\relax][c]{l's}
\cP^{\pi_a}_{E_A^{\wt{n}_a}\to E_A^{\wt{n}_a}},& $\pi_a\leq \Lambda_{\wt{n}_a}$\\
\id_{E_A^{\wt{n}_a}\to E_A^{\wt{n}_a}},& otherwise%
\end{IEEEeqnarraybox}\right.
\end{equation}

That is, if $\pi_a$ is no larger than $\Lambda_{\wt{n}_a}$, we use it to index the permutation channels from
$E_A^{\wt{n}_a}$ to $E_A^{\wt{n}_a}$ as usual; if $\pi_a$ is larger than $\Lambda_{\wt{n}_a}$, we set its
corresponding permutation channel to be the identity channel. Our conditional probability distribution
construction~\eqref{eq:conditional-probability} guarantees that this convention does not affect our result, as we will
show.

For each sequence $x^n\in\cX^n$, define the classical-quantum state:
\begin{equation}
  \omega^{x^n}_{\bm{\Lambda}U^nB^nE_B^n}
:=  \sum_{\bm{\pi},u^n}\frac{p_{\bm{\Lambda}\vert X^n}(\bm{\pi}\vert x^n)}{d^n}
    \proj{\bm{\pi}}_{\bm{\Lambda}}\ox\proj{u^n}_{U^n}\ox
    \cN^{\ox n}\circ\cE^{x^n}\circ\cP^{\bm{\pi}\vert x^n}\circ\cZ^{u^n}(\varphi^{\ox n}_{E_AE_B}).
\end{equation}
The constructions of the conditional probability distribution $p_{\bm{\Lambda}\vert X^n}(\bm{\pi}\vert x^n)$ 
in~\eqref{eq:conditional-probability} and permutation operation $\cP^{\bm{\pi}\vert x^n}$ 
in~\eqref{eq:permutation-construction} together yield
\begin{IEEEeqnarray}{rCl}
&~&  \omega^{x^n}_{\bm{\Lambda}U^nB^nE_B^n} \\
&=&  \sum_{\bm{\pi},u^n}\frac{p_{\bm{\Lambda}\vert X^n}(\bm{\pi}\vert x^n)}{d^n}
      \proj{\bm{\pi}}_{\bm{\Lambda}}\ox\proj{u^n}_{U^n}\ox
      \cN^{\ox n}\circ\cE^{x^n}\circ\cP^{\bm{\pi}\vert x^n}\circ\cZ^{u^n}(\varphi^{\ox n}_{E_AE_B}) \\
&=&  \bigotimes_{a\in\cX}\left(
        \frac{1}{\Lambda_{\wt{n}_a}d^{\wt{n}_a}}
        \sum_{\pi_a,u^{\wt{n}_a}}
        \proj{\pi_a}_{\Lambda_{\wt{n}_a}}\ox\proj{u^{\wt{n}_a}}_{U^{\wt{n}_a}}\ox
        \left(\cN\ox\cE^a\right)^{\ox \wt{n}_a} \ox
        \cP^{\pi_a}_{E_A^{\wt{n}_a}\to E_A^{\wt{n}_a}}\circ\cZ^{u^{\wt{n}_a}}
        (\varphi^{\ox \wt{n}_a}_{E_AE_B})\right) \\
&\equiv& \bigotimes_{a\in\cX}\omega^{a\vert t}_{\Lambda_{\wt{n}_a}U^{\wt{n}_a}B^{\wt{n}_a}E_B^{\wt{n}_a}},
  \label{eq:omega-a-x-n}
\end{IEEEeqnarray}
where $t$ is the type of $x^n$. We use $t$ instead of $x^n$ to indicate the fact that for all sequences $x^n$ of the
same type $t$, the conditional state $\omega^{x^n}_{\bm{\Lambda}U^nB^nE_B^n}$ necessarily reduces
to~\eqref{eq:omega-a-x-n}, which is only type dependent. One can check that each conditional state $\omega^{a\vert t}$
is of the form~\eqref{eq:lemma-state} defined in Lemma~\ref{lemma:permutation-existence} with $\cM\equiv\cN\ox\cE^a$.
This observation is essential in the achievability part. Consider the classical-quantum state induced by
$P_{X^n}(x^n)$:
\begin{equation}
  \omega_{\bm{\Lambda}X^nU^nB^nE_B^n} 
:= \sum_{x^n}p_{X^n}(x^n)\proj{x^n}_{X^n}\ox\omega^{x^n}_{\bm{\Lambda}U^nB^nE_B^n}.
\end{equation}
The reduced state of $\omega_{\bm{\Lambda}X^nU^nB^nE_B^n}$ satisfies:
\begin{IEEEeqnarray}{rCl}
  \omega_{X^nB^n}
&=&  \sum_{\bm{\pi},x^n,u^n}\frac{p_{\bm{\Lambda},X^n}(\bm{\pi}, x^n)}{d^n}\proj{x^n}_{X^n}\ox
      \cN^{\ox n}\circ\cE^{x^n}\circ\cP^{\bm{\pi}\vert x^n}\circ\cZ^{u^n}(\varphi^{\ox n}_{E_A}) \\
&=&  \sum_{x^n}p_{X^n}(x^n)\proj{x^n}_{X^n}\ox
      \cN^{\ox n}\circ \cE^{x^n}(\rho^{\ox n}_{E_A}) \\
&=&  \sigma_{XB}^{\ox n},
\end{IEEEeqnarray}
where the second equality follows as $\cZ^u(\varphi_A) = \varphi_A$ and $\varphi^{\ox n}_{E_A}$ is permutation
invariant. By the data-processing inequality of quantum mutual information, it holds that
\begin{equation}\label{eq:tmp-bound}
\Mutual(X^n{:}B^nE_B^n)_\omega \geq \Mutual(X^n{:}B^n)_\omega = \Mutual(X^n{:}B^n)_\sigma.
\end{equation}
Consider now the following chain of inequalities:
\begin{IEEEeqnarray}{rCl}
&~&   n\Mutual(XE_B{:}B)_\sigma - \Mutual(\bm{\Lambda}X^nU^n{:}B^nE_B^n)_\omega \\
&=&   \Mutual(X^nE_B^n{:}B^n)_\sigma - \Mutual(\bm{\Lambda}X^nU^n{:}B^nE_B^n)_\omega \\
&=&   \Mutual(E_B^n{:}B^n{\vert}X^n)_\sigma - \Mutual(\bm{\Lambda}U^n{:}B^nE_B^n{\vert}X^n)_\omega
      + \Mutual(X^n{:}B^n)_\sigma - \Mutual(X^n{:}B^nE_B^n)_\omega \\
&\leq&  \Mutual(E_B^n{:}B^n{\vert}X^n)_\sigma - \Mutual(\bm{\Lambda}U^n{:}B^nE_B^n{\vert}X^n)_\omega
          \label{eq:thm7-tmp3}\\
&=&   \sum_{x^n}p_{X^n}(x^n)
      \left[\Mutual(E_B^n{:}B^n)_{\sigma^{x^n}} - \Mutual(\bm{\Lambda}U^n{:}B^nE_B^n)_{\omega^{x^n}}\right] \\
&=&   \sum_{t\in\cT} \sum_{x^n\in T_t^{X^n}} p_{X^n}(x^n)
      \left[\Mutual(E_B^n{:}B^n)_{\sigma^{x^n}} - \Mutual(\bm{\Lambda}U^n{:}B^nE_B^n)_{\omega^{x^n}}\right] \\ 
&=&   \sum_{t\in\cT} d_tp_t
      \sum_{a\in\cX}\left[nt(a)\Mutual\left(E_B{:}B\right)_{\cN\circ\cE^a(\varphi)} 
      - \Mutual\left(\Pi_{nt(a)}U^{nt(a)}{:}B^{nt(a)}E_B^{nt(a)}\right)_{\omega^{a\vert t}}\right]
      \label{eq:thm7-tmp4} \\
&\leq& \sum_{t\in\cT} d_tp_t\sum_{a\in\cX} d\log(nt(a)+1)\label{eq:thm7-tmp1} \\
&\leq& \vert\cX\vert d\log(n+1),\label{eq:thm7-tmp2}
\end{IEEEeqnarray}
where~\eqref{eq:thm7-tmp3} follows from~\eqref{eq:tmp-bound},~\eqref{eq:thm7-tmp4} follows from~\eqref{eq:omega-a-x-n},
and~\eqref{eq:thm7-tmp1} follows from Lemma~\ref{lemma:permutation-existence}. Thus
\begin{IEEEeqnarray}{rCl}
  \Mutual_\varphi(\cN) 
&=& \Mutual(XE_B{:}B)_\sigma \\ 
&\leq&   \liminf_{n\to\infty}\frac{1}{n}\left\{\Mutual(\bm{\Lambda}X^nU^n{:}B^nE_B^n)_\omega 
        + \vert\cX\vert d\log(n+1)\right\} \\
&=& \liminf_{n\to\infty}\frac{1}{n}\Mutual(\bm{\Lambda}X^nU^n{:}B^nE_B^n)_\omega \\
&\leq& \ol{\chi}^{\pi}_{\varphi}\left(\cN\right).
\end{IEEEeqnarray}
We are done.
\end{IEEEproof}

\subsection{Strong converse}\label{sec:converse}

This section aims to prove Lemma~\ref{lemma:converse}. More concretely, we will show that $\Mutual_\rho(\cN)$ is a
strong converse bound for arbitrary $\rho_{E_AE_B}$-assisted classical communication under semi-global operations, even
when $\rho_{E_AE_B}$ is noisy. That is, the error probability necessarily converges to one in the limit of many channel
uses whenever the communication rate exceeds $\Mutual_\rho(\cN)$. Together with the achievability statement in
Lemma~\ref{lemma:achievability} for pure states $\varphi_{E_AE_B}$, we conclude that $\Mutual_\varphi(\cN)$ is a very
sharp dividing line between which communication rates are either achievable or unachievable asymptotically.

Our strong converse proof makes use of a meta-converse technique originally invented in~\cite{ogawa1999strong} and
further investigated in~\cite[Section 4.6]{hayashi2016quantum} (see also~\cite{sharma2013fundamental,wilde2014strong}
for more applications of this technique). Roughly speaking, the meta-converse states that a quantum divergence
satisfying some reasonable properties induces an upper bound on the success probability of any channel coding scheme.
Here we adopt the sandwiched \Renyi divergence
$\SRRel{\alpha}$~\cite{wilde2014strong,mueller-lennert2013quantum,frank2013monotonicity,beigi2013sandwiched}, which
meets all required properties. For two real numbers $x,y\in[0,1]$, denote the binary divergence
\begin{equation}
  \SRRel{\alpha}(x\Vert y) := \SRRel{\alpha}\left(x\proj{0}+(1-x)\proj{1}\rel y\proj{0}+(1-y)\proj{1}\right).
\end{equation}
Adapting the meta-converse argument into our communication scenario, we conclude the following relation for arbitrary
$\alpha\in(1,\infty)$ and blocklength $n$:
\begin{equation}\label{eq:sharma-warsi-meta}
      \SRRel{\alpha}\left(e(n,R)\rel1-2^{-nR}\right)\leq\max_{\omega}
      \SRRel{\alpha}\left(\omega_{XB^nE_B^n}\rel\omega_X\ox\left(\sigma_B^\star\ox\rho_{E_B}\right)^{\ox n}\right),
\end{equation}
where $R$ is the rate of communication, $e(n,R)$ is the error probability, $\omega_{XB^nE_B^n}$ is defined 
in~\eqref{eq:omega-XBnEBn}, and $\sigma_B^\star$ is a state achieving $\Mutual_\rho(\cN)$ 
w.r.t.~\eqref{eq:mutual-rel}. Let $s(n,R):=1-e(n,R)$ be the success probability. Evaluating the binary divergence
gives~\cite[(17)]{wilde2014strong}
\begin{IEEEeqnarray}{rCl}
  \SRRel{\alpha}\left(e(n,R)\rel1-2^{nR}\right)
&=& \frac{1}{\alpha-1}\log\left((e(n,R))^{\alpha}(1-2^{nR})^{1-\alpha}
                            + (s(n,R))^{\alpha}(2^{nR})^{1-\alpha}\right) \\
&\geq& \frac{1}{\alpha-1}\log\left((s(n,R))^{\alpha}(2^{nR})^{1-\alpha}\right) \\
&=& \frac{\alpha}{\alpha-1}\log s(n,R) + nR.
\end{IEEEeqnarray}
Substituting this into the meta converse~\eqref{eq:sharma-warsi-meta}, we get the following upper bound
on the success probability:
\begin{equation}\label{eq:succ-sand}
  s(n,R) \leq 2^{-n\frac{\alpha-1}{\alpha}\left(R - 
  \frac{1}{n}\max_{\omega}
  \SRRel{\alpha}\left(\omega_{XB^nE_B^n}\rel\omega_X\ox\left(\sigma_B^\star\ox\rho_{E_B}\right)^{\ox n}\right)
  \right)}.
\end{equation}
Consider now the following chain of inequalities:
\begin{IEEEeqnarray}{rCl}
&~&  \frac{1}{n}\max_{\omega}
      \SRRel{\alpha}\left(\omega_{XB^nE_B^n} \rel \omega_X\ox(\sigma_B^\star\ox\rho_{E_B})^{\ox n}\right) \\
&=&  \frac{1}{n}\max\sum_xp_x
      \SRRel{\alpha}\left(\omega^x_{B^nE_B^n} \rel (\sigma_B^\star\ox\rho_{E_B})^{\ox n}\right)
      \label{eq:cv-tmp1} \\
&\leq& \frac{1}{n}\max_x
        \SRRel{\alpha}\left(\omega^x_{B^nE_B^n} \rel (\sigma_B^\star\ox\rho_{E_B})^{\ox n}\right)
        \label{eq:cv-tmp2} \\
&=&    \frac{1}{n}\max_x\SRRel{\alpha}\left(
    \cN^{\ox n}\circ\cP^x\circ\bigotimes_{i=1}^n\cE^{[i]\vert x}\left(\rho^{\ox n}\right)
    \rel (\sigma_B^\star\ox\rho_{E_B})^{\ox n}\right) \\
&=&    \frac{1}{n}\max_x\SRRel{\alpha}\left(
    \cN^{\ox n}\left(\bigotimes_{i=1}^n\cE^{[i]\vert x}\left(\rho^{\ox n}\right)\right)
    \rel (\sigma_B^\star\ox\rho_{E_B})^{\ox n}\right)
    \label{eq:cv-tmp3} \\
&=&    \frac{1}{n}\max_x\SRRel{\alpha}\left(
        \bigotimes_{i=1}^n\left(\cN\circ\cE^{[i]\vert x}(\rho)\right)
        \rel(\sigma_B^\star\ox\rho_{E_B})^{\ox n}\right) \\
&=&    \frac{1}{n}\max_{x}\sum_{i=1}^n
        \SRRel{\alpha}\left(\cN\circ\cE^{[i]\vert x}(\rho)\rel\sigma_B^\star\ox\rho_{E_B}\right)
        \label{eq:cv-tmp4} \\
&\leq& \max_{\cE}\SRRel{\alpha}\left(\cN\circ\cE(\rho)\rel\sigma_B^\star\ox\rho_{E_B}\right),
\end{IEEEeqnarray}
where~\eqref{eq:cv-tmp1} follows from the direct-sum property,~\eqref{eq:cv-tmp3} follows from the fact that
$\cN^{\ox n}$ commutes with $\cP$ and $(\sigma_B^\star\ox\rho_{E_B})^{\ox n}$ is invariant under permutation,
and~\eqref{eq:cv-tmp4} follows from the additivity property w.r.t. tensor product.
Taking the limit $\alpha\to1$ on both sides of the above inequality gives
\begin{IEEEeqnarray}{rCl}
&~&  \lim_{\alpha\to1}\frac{1}{n}\max_{\omega}
      \SRRel{\alpha}\left(\omega_{XB^nE_B^n} \rel \omega_X\ox(\sigma_B^\star\ox\rho_{E_B})^{\ox n}\right) \\
&\leq&  \lim_{\alpha\to1}
      \max_{\cE}\SRRel{\alpha}\left(\cN\circ\cE(\rho)\rel\sigma_B^\star\ox\rho_{E_B}\right) \\
&=& \max_{\cE}\Rel\left(\cN\circ\cE(\rho)\rel\sigma_B^\star\ox\rho_{E_B}\right)\label{eq:limit-tmp} \\
&=& \Mutual_\rho(\cN),\label{eq:limit}
\end{IEEEeqnarray}
where~\eqref{eq:limit-tmp} follows from $\lim_{\alpha\to1}\SRRel{\alpha}=\Rel$ and~\eqref{eq:limit} follows
from that $\sigma_B^\star$ is a state achieving $\Mutual_\rho(\cN)$. When $R > \Mutual_\rho(\cN)$,
\eqref{eq:succ-sand} and~\eqref{eq:limit} together guarantee that there exists some $\alpha>1$ for which the exponent
\begin{equation}
  \frac{\alpha-1}{\alpha}\left(R - \frac{1}{n}\max_{\omega}
  \SRRel{\alpha}\left(\omega_{XB^nE_B^n}\rel\omega_X\ox\left(\sigma_B^\star\ox\rho_{E_B}\right)^{\ox n}\right)
  \right)
\end{equation}
is strictly positive, which implies the success probability decays exponentially fast to $0$. This concludes the strong
converse part.

\subsection{Comparison with previous results}\label{sec:comparison}

Assume $E_A\cong A$ and let $\gamma_{BE_B} := \cN_{A\to B}(\varphi_{E_AE_B})$. The achievability part of the BSST
theorem~\cite{bennett1999entanglement,bennett2002entanglement} showed that $\Mutual(B{:}E_B)_\gamma$ is an achievable
rate for $\varphi$-assisted classical communications. The constructed protocol used global encoding operations. Later,
Shor~\cite{shor2004classical} proposed a new protocol using semi-global operations to achieve
$\Mutual(B{:}E_B)_\gamma$. Surprisingly, we find that $\Mutual_\varphi(\cN)$ is larger than $\Mutual(B{:}E_B)_\gamma$.
That is to say, we find a larger achievable rate for the $\varphi$-assisted classical communication, when semi-global
operations are allowed.

\begin{proposition}\label{eq:mul-lower-bound}
It holds that $\Mutual_\varphi(\cN) \geq \Mutual(B{:}E_B)_\gamma$.
\end{proposition}
\begin{IEEEproof}
Recall the min-max formula of $\Mutual_\varphi(\cN)$ in Proposition~\ref{prop:rel-characterization}, we have
\begin{IEEEeqnarray}{rCl}
  \Mutual_\varphi(\cN) 
=&& \min_{\tau_B}\max_{\cE_{E_A\to A}}
  \Rel\left(\cN_{A\to B}\circ\cE_{E_A\to A}(\varphi_{E_AE_B})\rel\tau_B\ox\varphi_{E_B}\right)
  \label{eq:min-max-2-max-min} \\
&\geq& \max_{\cE_{E_A\to A}}\min_{\tau_B}
  \Rel\left(\cN_{A\to B}\circ\cE_{E_A\to A}(\varphi_{E_AE_B})\rel\tau_B\ox\varphi_{E_B}\right) \\
=&& \max_{\cE_{E_A\to A}}\Mutual(B{:}E_B)_{\cN\circ\cE(\varphi)} \\
&\geq& \Mutual(B{:}E_B)_\gamma,\label{eq:use-id}
\end{IEEEeqnarray}
where~\eqref{eq:min-max-2-max-min} follows from the fact that $\min\max$ is no less than $\max\min$, 
and~\eqref{eq:use-id} by choosing $\cE$ to be the identity channel.
\end{IEEEproof}

\section{Examples}\label{sec:examples}

\subsection{Pinching channels}\label{sec:pinching}

Proposition~\ref{prop:chi-mutual-ea} states that $\chi_\rho(\cN)\leq\Mutual_\rho(\cN)$ holds in general. In this
section, we show that for pinching channels this inequality can be strict. That is, there exist a pinching channel
$\cP$ and a pure state $\varphi_ {E_AE_B}$ for which $\chi_\varphi(\cP) < \Mutual_\varphi(\cP)$. In the light of
Proposition~\ref{prop:capacity-product} and Theorem~\ref{thm:capacity}, this result witnesses the power of permutation
in classical communication -- the $\varphi$-assisted classical capacity of $\cP$ using local operations and global
permutations (aka. semi-global operations) is strictly larger than the $\varphi$-assisted classical capacity of $\cP$
using local operations only.

Let $A_1,\cdots, A_k$ be $k$ Hilbert spaces that each is $d_i$-dimensional. Let $\bm{A}=\oplus_{i=1}^kA_i$ be the
direct sum of these spaces. By definition, the dimension of $\bm{A}$ is $d=\sum_{i=1}^kd_i$. Let $\Pi_i$ be the
projection onto $A_i$ w.r.t. $\bm{A}$. The pinching channel $\cP_{\bm{A}\to \bm{A}}$ is defined as
\begin{equation}
  \cP_{\bm{A}\to \bm{A}}(\rho) = \sum_{i=1}^{k}\Pi_i\rho\Pi_i.
\end{equation}
This channel is a special case of phase-damping channel that removes off-diagonal blocks of the input matrix. Since
$\Mutual(\cP\vert\rho_{\bm{A}})$ is concave in $\rho_{\bm{A}}$~\cite[(8.45)]{hayashi2016quantum}, $\Mutual(\cP)$ is
achieved among states $\rho_{\bm{A}}$ of the form
\begin{equation}
    \rho_{\bm{A}} = \sum_{i=1}^kp_i\frac{\Pi_i}{d_i},
\end{equation}
where $\bm{p}\equiv(p_1,\cdots,p_k)$ forms a probability distribution. Let $\varphi_{\bm{A}'\bm{A}}$ be a purification
of $\rho_{\bm{A}}$ and set $\sigma_{\bm{A}'\bm{A}}=\cP_{\bm{A}\to\bm{A}}(\varphi_{\bm{A}'\bm{A}})$. By definition,
$\cP(\rho_{\bm{A}})=\rho_{\bm{A}}$ and thus $\sigma_{\bm{A}'}=\sigma_{\bm{A}}$. Then
\begin{IEEEeqnarray}{rCl}
  \Mutual(\cP) 
&=& \max_{\bm{p}}\Mutual(\bm{A}'{:}\bm{A})_{\sigma} \\ 
&=& \max_{\bm{p}}\left\{\Shannon(\bm{A}')_\sigma + \Shannon(\bm{A})_\sigma
        - \Shannon(\bm{A}'\bm{A})_\sigma\right\} \\
&=& \max_{\bm{p}}\left\{2\left(\Shannon(\bm{p}) + \sum_{i=1}^kp_i\log d_i\right) - \Shannon(\bm{p})\right\} \\
&=& \log\Delta - \min_{\bm{p}}\Rel\left(\bm{p}\rel(d_i^2/\Delta)\right),\label{eq:pinching-1}
\end{IEEEeqnarray}
where $\Delta\equiv\sum_{i=1}^kd_i^2$ and $(d_i^2/\Delta)\equiv(d_1^2/\Delta,\cdots,d_k^2/\Delta)$ denotes a
probability distribution. Since quantum relative entropy is non-negative, the minimization in~\eqref{eq:pinching-1} is
achieved when $\bm{p}^\star=(d_i^2/\Delta)$, that is, $p^\star_i=d_i^2/\Delta$, and the corresponding optimal state
$\rho^\star_{\bm{A}}$ has the form
\begin{IEEEeqnarray}{rCl}
    \rho^\star_{\bm{A}} &=& \sum_{i=1}^kp^\star_i\frac{\Pi_i}{d_i} = \sum_{i=1}^k\frac{d_i}{\Delta}\Pi_i.
\end{IEEEeqnarray}

Assume now that there exist indices $i\neq j$ for which $d_i\neq d_j$. Under this assumption, $\rho^\star_{\bm{A}}$ is
not completely mixed on its support since $\bm{p}^\star$ is not uniform. Let $\varphi^\star_{\bm{A}'\bm{A}}$ be a
purification of $\rho^\star_{\bm{A}}$. In the light of Proposition~\ref{prop:chi-mutual-equal-condition}, we conclude
that for the $\varphi^\star_{\bm{A}'\bm{A}}$-assisted classical communication over $\cP$, permutation \textit{does}
improve the communication rate compared to the case when only local encoding is allowed, as captured in the following
proposition, whose proof can be found in Appendix~\ref{appx:prop:pinching-bound}.
\begin{proposition}\label{prop:pinching-bound}
The strict inequality
$C^{\ox,\varepsilon}_{\varphi^\star}(\cP) < C^{\pi,\varepsilon}_{\varphi^\star}(\cP)=\log\Delta$ 
holds for $\varepsilon\in[0,1)$.
\end{proposition}

Inspecting the proof for Proposition~\ref{prop:pinching-bound}, we obtain an upper bound on the gap between
$C^{\ox,\varepsilon}_{\varphi^\star}(\cP)$ and $C^{\pi,\varepsilon}_{\varphi^\star}(\cP)$.
\begin{corollary}
The inequality
$C^{\pi,\varepsilon}_{\varphi^\star}(\cP) - C^{\ox,\varepsilon}_{\varphi^\star}(\cP) \leq\Shannon(\bm{p}^\star)$ 
holds for $\varepsilon\in[0,1)$.
\end{corollary}

\subsection{Covariant channels}\label{sec:covariant}

In this section we investigate the equality condition for $\chi_\rho(\cN)\leq\Mutual_\rho(\cN)$. More concretely, we
show that for the class of covariant channels, these two information measures are equal for arbitrary state
$\rho_{E_AE_B}$.

Consider the (projective) representations $f_A$ and $f_B$ of a compact group $G$ on $\cH_A$ and $\cH_B$, respectively,
such that $f_A$ is irreducible. We call a quantum channel $\cN_{A\to B}$ covariant with respect to $\{f_A(g),
f_B(g)\}_{g\in G}$, if
\begin{equation}
  \cN_{A \to B}(f_A(g) (\cdot) f_A(g)^\dagger) = f_B(g){\cal N}_{A \to B}(\cdot)f_B(g)^\dagger
\end{equation}
for all $g\in G$. As examples, erasure channel~\cite[Example 5.12]{hayashi2016quantum} and depolarizing
channel~\cite[Example 5.3]{hayashi2016quantum} are covariant when $G$ is the unitaries on the input system. Generalized
Pauli channel~\cite[Example 5.8]{hayashi2016quantum} is covariant when $G$ is the discrete Weyl representation. The
qubit phase damping channel~\cite[Example 5.10]{hayashi2016quantum} is covariant when $G$ is the discrete Weyl
representation. However, general qudit phase damping channel is not necessarily covariant.

For a covariant channel $\cN$, we have the following known facts~\cite[Section 9.7.1]{hayashi2016quantum}:
\begin{IEEEeqnarray}{rCl}
  \chi(\cN) &=& \Shannon(\cN_{A\to B}(\pi_A))  - \min_{\rho_A}\Shannon(\cN_{A\to B}(\rho_A)),
      \label{eq:depolarizing-chi} \\
  \Mutual(\cN) &=& \Shannon(\cN_{A\to B}(\pi_A)) + \log d_A - \Shannon(\cN_{A\to B}(\Phi_{A'A})),
\end{IEEEeqnarray}
where $\ket{\Phi_{A'A}}:=\sum_{i=1}^{d_A}\sqrt{1/d_A}\ket{ii}$ is the maximally entangled state of rank $d_A$.
Furthermore, we show that $\chi_\rho(\cN)$ is equal to $\Mutual_\rho(\cN)$ and obtain an useful expression for these
quantities. The proof is given in Appendix~\ref{appx:prop:covariant}.

\begin{proposition}\label{prop:covariant}
Let $\rho_{E_AE_B}$ be a bipartite state and $\cN_{A\to B}$ be a covariant channel. It holds that
\begin{equation}\label{eq:covariant}
\chi_\rho(\cN) =\Mutual_\rho(\cN) 
= \Shannon(\cN_{A\to B}(\pi_A)) + \Shannon(\rho_{E_B}) 
- \min_{\cE_{E_A \to A}} \Shannon(\cN_{A\to B}\circ\cE_{E_A \to A}(\rho_{E_AE_B})).
\end{equation}
\end{proposition}
\begin{remark}
We emphasize that~\eqref{eq:covariant} holds even when $\rho_{E_AE_B}$ is not pure. In the light of
Proposition~\ref{prop:capacity-product} and Lemma~\ref{lemma:converse}, \eqref{eq:covariant} implies that the
transmission rate $\chi_\rho(\cN)$ is optimal among semi-global encoding, whenever the channel is covariant.
\end{remark}
\begin{remark}
For a covariant channel $\cN$,~\eqref{eq:upper-bound} becomes
\begin{equation}
\chi_\rho(\cN) - \chi(\cN) = \Mutual_\rho(\cN) - \chi(\cN) \leq D_F(\rho_{E_AE_B}).
\end{equation}
Substituting~\eqref{eq:depolarizing-chi} and~\eqref{eq:covariant}, we reach the following non-trivial lower bound which
might be of independent interests regarding covariant channels:
\begin{equation}
   \min_{\cE_{E_A \to A}} \Shannon(\cN_{A\to B}\circ\cE_{E_A \to A}(\rho_{E_AE_B}))
\geq  \min_{\rho_A}\Shannon(\cN_{A\to B}(\rho_A)) + \Shannon(\rho_{E_B}) - D_F(\rho_{E_AE_B}).
\end{equation} 
\end{remark}

\subsection{Erasure channels}

As a concrete example of covariant channels discussed above, we consider the qudit erasure channel, whose corresponding
group $G$ is the unitaries on the input system. Specifically, the qudit erasure channel is defined as
\begin{equation}
  \cE_{d,p}(\rho) = (1 - p)\rho + p\proj{e},
\end{equation}
where $p\in[0,1]$ and $\ket{e}$ is an erasure symbol orthogonal to the qudit space. It holds that~\cite[Section
9.7.6]{hayashi2016quantum}:
\begin{IEEEeqnarray}{rCl}
  \chi(\cE_{d,p}) &=& (1 - p)\log d, \\
  \Mutual(\cE_{d,p}) &=& 2(1 - p)\log d.
\end{IEEEeqnarray}

We assume the following two-qudit pure entangled state is available
\begin{equation}
  \ket{\Phi_{\bm{\lambda}}} := \sum_{i=1}^d\sqrt{\lambda_i}\ket{ii},
\end{equation}
where $\bm{\lambda}\equiv(\lambda_1,\cdots,\lambda_d)$ satisfying $\lambda_i\geq0$ and $\sum_i\lambda_i=1$. Set
$\Phi_{\bm{\lambda}}\equiv\proj{\Phi_{\bm{\lambda}}}$ and let $\Shannon(\bm{\lambda})$ denote the entropy of
$\bm{\lambda}$. When $d=2$ (the two-qubit case), we write for simplicity
$\Phi_{\lambda}\equiv\Phi_{(\lambda,1-\lambda)}$, where $\lambda\in[0,1/2]$. As shown in
Appendix~\ref{appx:prop:erasure-channel}, the $\Phi_{\bm{\lambda}}$-assisted classical capacities of $\cE_{d,p}$ using
product encoding and semi-global encoding have analytic expression.
\begin{proposition}\label{prop:erasure-channel}
Let $\varepsilon\in[0,1)$. It holds that
\begin{equation}
    \chi_{\Phi_{\bm{\lambda}}}(\cE_{d,p})
=   C^{\ox,\varepsilon}_{\Phi_{\bm{\lambda}}}(\cE_{d,p})
=   C^{\pi,\varepsilon}_{\Phi_{\bm{\lambda}}}(\cE_{d,p}) 
=   \Mutual_{\Phi_{\bm{\lambda}}}(\cE_{d,p})
=   (1-p)(\log d +\Shannon(\bm{\lambda})).
\end{equation}
\end{proposition}

\begin{figure}[htbp]
  \centering
  \includegraphics[width=0.6\textwidth]{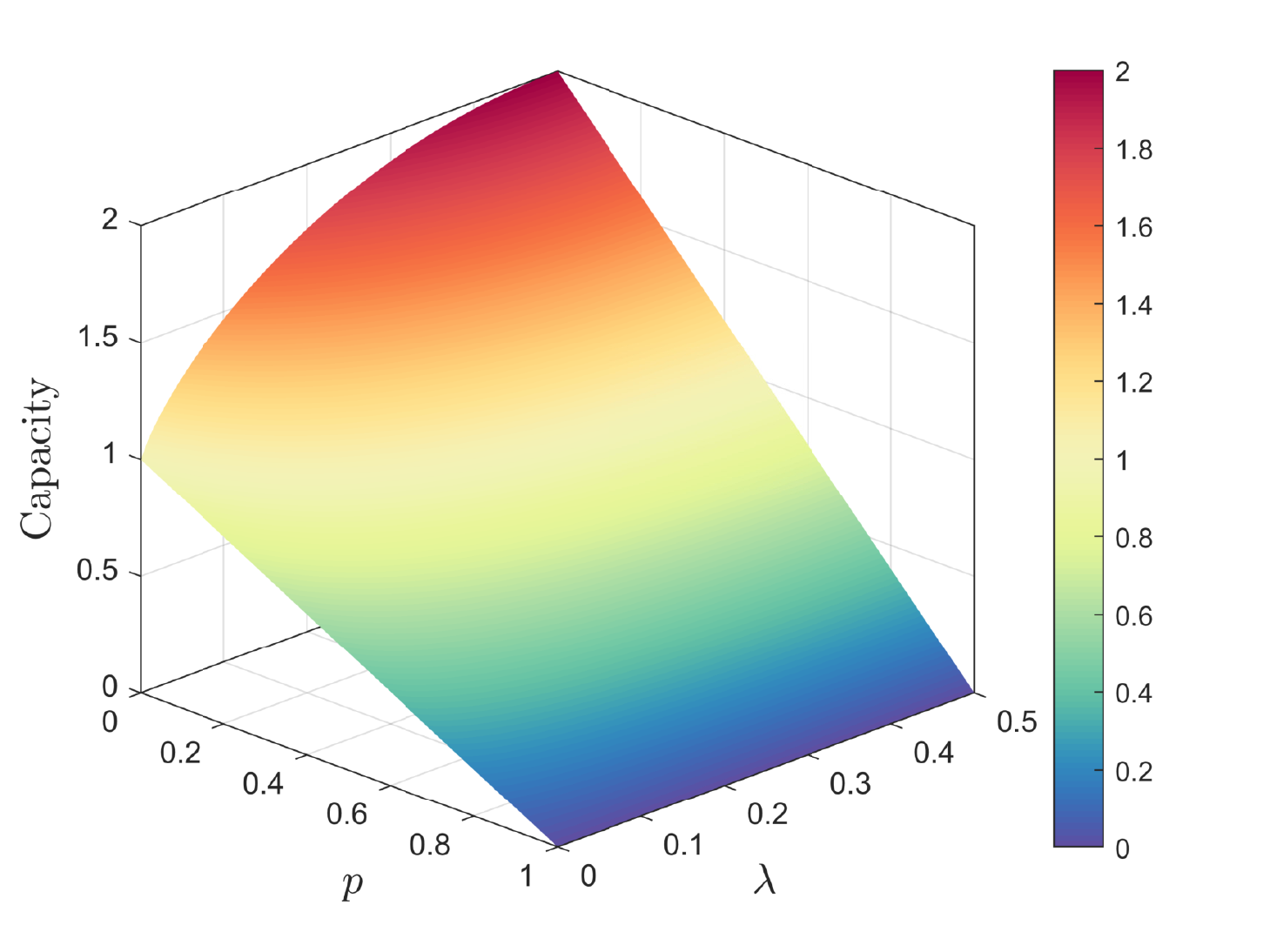}
  \caption{The $\Phi_{\lambda}$-assisted classical capacity of the erasure channel $\cE_{2,p}$ using semi-global
      operations as a function of the erasure parameter $p$ and the state parameter $\lambda$. When $\lambda=0$, we
      recover $\chi(\cE_{2,p})$; when $\lambda=1/2$, we recover $\Mutual(\cE_{2,p})$.}
  \label{fig:erasure-channel}
\end{figure}

Note that $\chi(\cE_{d,p})$ is recovered when $\Phi_{\bm{\lambda}}$ is product, i.e., $\bm{\lambda}=(1,0,\cdots,0)$,
while $\Mutual(\cE_{d,p})$ is recovered when $\Phi_{\bm{\lambda}}$ is maximally entangled, i.e.,
$\bm{\lambda}=(1/d,\cdots,1/d)$. As an illustrative example, Fig.~\ref{fig:erasure-channel} shows how the
$\Phi_\lambda$-assisted capacity varies with parameters $p$ and $\lambda$ for the qubit erasure channel $\cE_{2,p}$ and
two-qubit pure entangled state $\Phi_{\lambda}$.

\begin{figure}[htbp]
  \centering
  \includegraphics[width=0.6\textwidth]{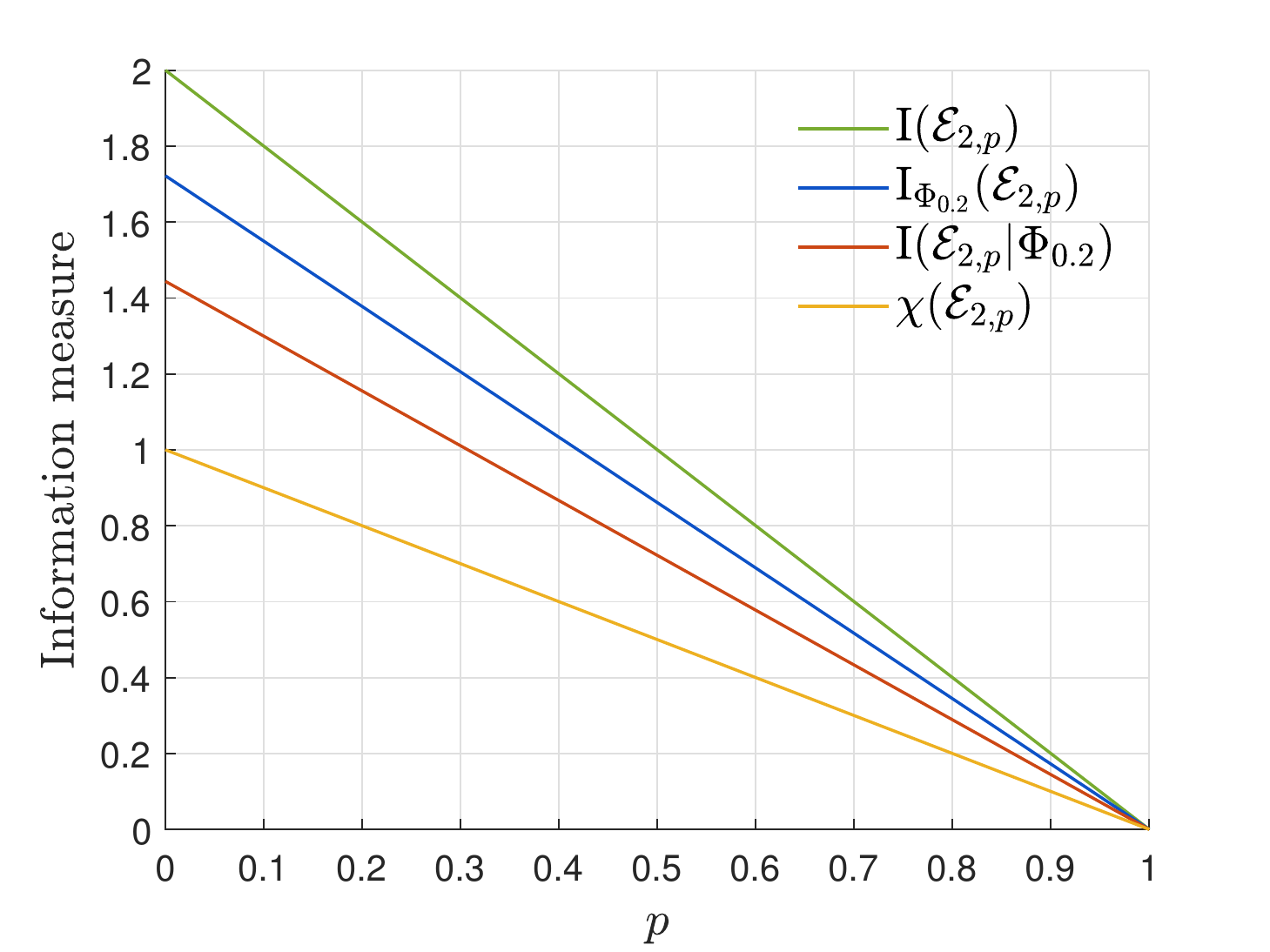}
  \caption{Comparison among various information measures of the qubit erasure channel $\cE_{2,p}$: $\chi(\cE_{2,p})$,
          $\Mutual(\cE_{2,p}\vert\Phi_{0.2})$, $\Mutual_{\Phi_{0.2}}(\cE_{2,p})$, and $\Mutual(\cE_{2,p})$.}
  \label{fig:erasurek-channel-compare}
\end{figure}

We also compare the bounds discussed in Proposition~\ref{eq:mul-lower-bound} for the qubit erasure channel. By
Proposition~\ref{eq:mul-lower-bound}, we have the following chain of inequalities:
\begin{equation}
  \chi(\cE_{2,p}) \leq \Mutual(\cE_{2,p}\vert\Phi_\lambda) \leq \Mutual_{\Phi_\lambda}(\cE_{2,p}) 
  \leq \Mutual(\cE_{2,p}).
\end{equation}
Recall that $\Mutual(\cE_{2,p}\vert\Phi_\lambda)$ is defined in~\eqref{eq:mutual-wrt-rho}. Actually, for $\cE_p$
these inequalities can all be strict. In Fig.~\ref{fig:erasurek-channel-compare} we compare these quantities on the
full range $p\in[0,1]$ with fixed $\lambda=0.2$. The strict gap between $\Mutual(\cE_{2,p}\vert\Phi_{0.2})$ and
$\Mutual_{\Phi_{0.2}}(\cE_{2,p})$ for $p\in(0,1)$ indicate that our obtained capacity formula for the
$\varphi$-assisted classical communication, when only semi-global operations are allowed, is better than the achievable
rate previous derived in~\cite{shor2004classical}.

\section{Conclusion}\label{sec:conclusion}

We have investigated a special case of classical communication over quantum channels, in which the set of available
encoding is restricted to local operations and global permutations and multiple copies of an entangled state are
preshared among the sender and the receiver in product form. A capacity formula for the classical capacity was
established when the preshared state is pure. Furthermore, we showed that the capacity satisfies the strong converse
property and thus the capacity formula served as a sharp dividing line between achievable and unachievable rates of
communication. As demonstrative examples, we considered various quantum channels of interests and showed that their
classical capacities have analytical expression. Our result highlighted the importance of random permutation in
entanglement assisted classical communication -- it can enhance classical communication compared to the case when only
local encoding is allowed. As by-product, we introduced a new quantity $\Mutual_\rho(\cN)$ -- the $\rho$-assisted
mutual information of $\cN$ -- to quantify how much classical correlation Alice and Bob can establish by using the
$\cN$, under the assistance of a preshared $\rho$. We showed that the gap between $\Mutual_\rho(\cN)$ and the Holevo
capacity is upper bounded by the discord of formation of $\rho$.

An important open problem is whether our derived capacity formula can be extended to the noisy entanglement assistance
case, i.e., is $\Mutual_\rho(\cN)$ equal to $C^{\pi,\varepsilon}_\rho\left(\cN\right)$ for arbitrary bipartite quantum
state $\rho_{E_AE_B}$ and channel $\cN_{A\to B}$? It is also interesting to study how large the gap between
$\Mutual_\varphi(\cN)$ and $\chi_\varphi(\cN)$ can be. This gap quantitatively witnesses the power of random
permutations in classical communication.

\section*{Acknowledgment}
\addcontentsline{toc}{section}{Acknowledgment}

KW thanks Xin Wang for helpful discussion. MH was supported in part by Japan Society for the Promotion of Science
(JSPS) Grant-in-Aid for Scientific Research (A) No. 17H01280, (B) No. 16KT0017, and Kayamori Foundation of
Informational Science Advancement.



\appendices
\renewcommand{\theequation}{\thesection.\arabic{equation}}

\section{Proof of Proposition~\ref{prop:rel-characterization}}\label{appx:prop:rel-characterization}

\begin{IEEEproof}
\eqref{eq:hol-rel} was proved in~\cite[(19)]{schumacher2001optimal} and~\eqref{eq:mut-rel} was proved
in~\cite[(5)]{bennett2002entanglement}. We are going to prove~\eqref{eq:chi-ea-rel}.~\eqref{eq:mutual-rel} can
shown using the same technique. For arbitrary $p_X$ and $\sigma_{BE_B}$, define the quantity
\begin{equation}
  J(\cN,p_X,\sigma_{BE_B})
:= \Mutual\left(X{:}BE_B\right)_\omega + \Rel\left(\omega_{BE_B}\rel\sigma_{BE_B}\right).
\end{equation}
By definition, we have
\begin{IEEEeqnarray}{rCl}
    J(\cN,p_X,\sigma_{BE_B})
&:=& \Mutual\left(X{:}BE_B\right)_\omega + \Rel\left(\omega_{BE_B}\rel\sigma_{BE_B}\right) \\
&=& \Rel\left(\omega_{XBE_B}\rel\omega_X\ox\sigma_{BE_B}\right) \\
&=& \sum_xp_X(x)\Rel\left(\omega^x_{BE_B}\rel\sigma_{BE_B}\right),
\end{IEEEeqnarray}
where $\omega^x_{BE_B} := \cN_{A\to B}\circ\cE^x_{E_A\to A}(\rho_{E_AE_B})$ and the last equality follows from the
direct-sum property of quantum relative entropy. It then follows that $J(\cN,p_X,\sigma_{BE_B})$ is linear in $p_X$ and
convex in $\sigma_{BE_B}$. By the positivity of quantum relative entropy we have
\begin{equation}
  \min_{\sigma_{BE_B}}J(\cN,p_X,\sigma_{BE_B}) = \Mutual\left(X{:}BE_B\right)_\omega.
\end{equation}
Thus
\begin{IEEEeqnarray}{rCl}
    \chi_\rho(\cN) 
&=&  \max_{p_X}\Mutual\left(X{:}BE_B\right)_\omega \\
&=&  \max_{p_X}\min_{\sigma_{BE_B}}J(\cN,p_X,\sigma_{BE_B}) \\
&=&  \min_{\sigma_{BE_B}}\max_{p_X}J(\cN,p_X,\sigma_{BE_B})\label{eq:min-max} \\
&=&  \min_{\sigma_{BE_B}}\max_{p_X}\sum_xp_X(x)\Rel\left(\omega^x_{BE_B}\rel\sigma_{BE_B}\right) \\
&=&  \min_{\sigma_{BE_B}}\max_{\cE_{E_A\to A}}\Rel\left(
      \cN_{A\to B}\circ\cE_{E_A\to A}(\rho_{E_AE_B})\rel\sigma_{BE_B}\right),
\end{IEEEeqnarray}
where~\eqref{eq:min-max} follows from Sion's minimax theorem~\cite{sion1958general}.
\end{IEEEproof}

\section{Proof of Lemma~\ref{lemma:difference}}\label{appx:lemma:difference}

\begin{IEEEproof}
By the definition of $\omega_{XBE_B}$ (cf.~\eqref{eq:omega_XBEB}), $X$ and $E_B$ are independent. We have
\begin{IEEEeqnarray}{rCl}
&~&   \Mutual(XE_B{:}B)_\omega - \Mutual(X{:}BE_B)_\omega \\
&=&  \Shannon(X)_\omega + \Shannon(E_B)_\omega + \Shannon(B)_\omega - \Shannon(XBE_B)
- \Shannon(X)_\omega - \Shannon(BE_B)_\omega + \Shannon(XBE_B) \\
&=& \Shannon(E_B)_\omega + \Shannon(B)_\omega - \Shannon(BE_B)_\omega \\
&=& \Mutual(B{:}E_B)_\omega.
\end{IEEEeqnarray}
\end{IEEEproof}

\section{Proof of Proposition~\ref{prop:mutual-relation}}\label{appx:prop:mutual-relation}

The proof of Proposition~\ref{prop:mutual-relation} relies on the following two lemmas.
\begin{lemma}\label{L12}
The function $\Mutual(\cN_{A\to B}\vert\rho_A)$, defined in~\eqref{eq:mutual-wrt-rho}, is strictly concave in $\rho_A$.
\end{lemma}
\begin{IEEEproof}
This can be shown by applying the Petz's equality condition for the monotonicity of relative 
entropy~\cite[Corollary 6.1]{hayashi2016quantum} to~\cite[Exercise 8.24]{hayashi2016quantum}. 
For completeness, we write down the details.

Let $\rho_1$ and $\rho_2$ be arbitrary two quantum states such that $\rho_1\neq\rho_2$ and let $\lambda\in(0,1)$. We
now show the following strict inequality
\begin{align}
  \Mutual\left(\cN\middle\vert\lambda\rho_1 + (1-\lambda)\rho_2\right)
> \lambda\Mutual\left(\cN\middle\vert\rho_1\right)
+ (1-\lambda)\Mutual\left(\cN\middle\vert\rho_2\right),
\end{align}
from which the strict concavity property follows. Let $U_{A\to BE}$ be a Stinespring representation of $\cN_{A\to B}$.
Let $\ol{\rho} = \lambda\rho_1 + (1-\lambda)\rho_2$, $\ol{\sigma}_{BE} = U\ol{\rho}U^\dagger$, and
$\sigma^x_{BE}=U\rho_xU^\dagger$ for $x=1,2$. Then $\ol{\sigma}_{BE} = \lambda\sigma^1_{BE}+(1-\lambda)\sigma^2_{BE}$
and $\ol{\sigma}_E=\lambda\sigma^1_E+(1-\lambda)\sigma^2_E$. We need the following two statements~\cite[Theorem
7]{ruskai2002inequalities}:
\begin{enumerate}
  \item 
      $\Rel\left(\ol{\sigma}_{BE}\rel\1_B\ox\ol{\sigma}_E\right)
      = \lambda\Rel\left(\sigma^1_{BE}\rel\1_B\ox\sigma^1_E\right)
      + (1-\lambda)\Rel\left(\sigma^2_{BE}\rel\1_B\ox\sigma^2_E\right)$
      if and only if
      \begin{equation}
          \log\ol{\sigma}_{BE} - \log\ol{\sigma}_B 
        = \log\sigma^1_{BE} - \log\sigma^1_B 
        = \log\sigma^2_{BE} - \log\sigma^2_B.
      \end{equation}
  \item 
      $\Rel\left(\ol{\sigma}_B\rel\1_B\right)
      = \lambda\Rel\left(\sigma^1_B\rel\1_B\right)
      + (1-\lambda)\Rel\left(\sigma^2_B\rel\1_B\right)$
      if and only if $\log\ol{\sigma}_B = \log\sigma^1_B = \log\sigma^2_B$.
\end{enumerate}
Since $\sigma^1_{BE}=U\rho_1U^\dagger\neq U\rho_2U^\dagger=\sigma^2_{BE}$, the above two equalities regarding quantum
relative entropy cannot both hold. Using this fact, we have the following chain of inequalities:
\begin{IEEEeqnarray}{rCl}
  \Mutual(\cN\vert\ol{\rho}) 
&=& \Shannon(B\vert E)_{\ol{\sigma}} + \Shannon(B)_{\ol{\sigma}} \\
&=& - \Rel\left(\ol{\sigma}_{BE}\rel\1_B\ox\ol{\sigma}_E\right) - \Rel\left(\ol{\sigma}_B\rel\1_B\right) \\
&>& - \left[  \lambda\Rel\left(\sigma^1_{BE}\rel\1_B\ox\sigma^1_E\right)
            + (1-\lambda)\Rel\left(\sigma^2_{BE}\rel\1_B\ox\sigma^2_E\right)\right]\nonumber \\
&~& - \left[\lambda\Rel\left(\sigma^1_B\rel\1_B\right) + (1-\lambda)\Rel\left(\sigma^2_B\rel\1_B\right)\right] \\
&=& \lambda\left[\Shannon(B\vert E)_{\sigma^1} + \Shannon(B)_{\sigma^1}\right]
    + (1-\lambda)\left[\Shannon(B\vert E)_{\sigma^2} + \Shannon(B)_{\sigma^2}\right] \\
&=& \lambda\Mutual\left(\cN\middle\vert\rho_1\right) + (1-\lambda)\Mutual\left(\cN\middle\vert\rho_2\right).
\end{IEEEeqnarray}
We are done.
\end{IEEEproof}

\begin{lemma}\label{L13}
The maximum
\begin{align}\label{eq:L13}
\max_{\rho_{A'A}}\Rel\left(\cN_{A\to B}(\rho_{A'A})\rel\rho_{A'}\ox\tr_{A'}\cN_{A\to B}(\rho_{A'A})\right)
\end{align}
is attained only when $\rho_{A'A}$ can be converted to $\varphi^\star_{A'A}$ via a local unitary on $A'$.
\end{lemma}

\begin{IEEEproof}
Assume $\bar{\rho}_{A'A}$ achieves the maximum in~\eqref{eq:L13}. Let $\bar{\varphi}_{RA'A}$ be a purification of
$\bar{\rho}_{A'A}$. Applying the Petz's equality condition for the monotonicity of relative entropy \cite[Corollary
6.1]{hayashi2016quantum} to the following inequality
\begin{align}
  \Rel\left(\cN_{A\to B}(\bar{\rho}_{A'A})\rel
  \bar{\rho}_{A'}\ox \tr_{A'}\cN_{A\to B}(\bar{\rho}_{A'A})\right)
\leq   
  \Rel\left(\cN_{A\to B}(\bar{\varphi}_{RA'A})\rel
  \bar{\varphi}_{RA'}\ox\tr_{RA'}\cN_{A\to B}(\bar{\varphi}_{RA'A})\right),
\end{align}
we find that $\bar{\varphi}_{RA'}$ has the form $\bar{\varphi}_{R}\otimes\bar{\varphi}_{A'}$, which implies that
$\bar{\rho}_{A'A}$ is actually a pure state. Hence, combining Lemma \ref{L12}, we obtain the desired statement.
\end{IEEEproof}

\begin{IEEEproof}[Proof of Proposition~\ref{prop:mutual-relation}]
Inspecting~\eqref{eq:mut-rel} and~\eqref{eq:mutual-rel} and considering the case with $\varphi_{AA'}=\cE_{E_A\to
A}(\rho_{E_AE_B})$, we obtain the inequality $\Mutual_\rho(\cN)\leq\Mutual(\cN)$.

As for the necessary and sufficient condition, we focus on the RHS. of~\eqref{eq:mut-rel} and~\eqref{eq:mutual-rel}. We
find that $\Mutual_\rho(\cN)=\Mutual(\cN)$ holds if and only if there exists a channel $\cE_{E_A\to A}$ such that
$\varphi^\star_{AA'}=\cE_{E_A\to A}(\rho_{E_AE_B})$. By Lemma~\ref{L13}, this condition is equivalent to the condition
that $\varphi^\star_{AA'}=\rho_{E_AE_B}$ and $\cE_{E_A\to A}$ preserves the eigenspace of the reduced density
$\tr_{A'}\varphi^\star_{AA'}$ for a non-zero eigenvalue, and thus its action on the eigenspace composed of non-zero
eigenvalues is an unitary. Hence, we identify the equality condition.
\end{IEEEproof}

\section{Proof of Proposition~\ref{prop:chi-mutual-equal-condition}}\label{appx:prop:chi-mutual-equal-condition}

\begin{IEEEproof}
Due to Proposition~\ref{prop:chi-mutual-ea}, $\chi_\rho(\cN)=\Mutual_\rho(\cN)$ if and only if there exist a
distribution $p_X$ and a set of encoding operations $\{\cE^x_{E_A\to A}\}$ such that the corresponding
$\omega_{XBE_B}$, as constructed in~\eqref{eq:omega_XBEB}, satisfies the property that its reduced state
$\omega_{BE_B}$ has independent systems $E_B$ and $B$. On the other hand, from the proof of the equality condition for
$\Mutual_\rho(\cN)=\Mutual(\cN)$ (cf. Proposition~\ref{prop:mutual-relation}), we know each $\cE_{E_A\to A}^x$ is a
unitary that acts only on the support of the reduced state $\tr_{A'}\varphi^\star_{A'A}$ and it must hold that
$\sum_xp_x\cE^x_{E_A\to A}(\rho_{E_A})=\tr_{A'}\varphi^\star_{A'A}$. Under these constraints, $p_X$ and
$\{\cE^x_{E_A\to A}\}$ exist only when the $\tr_{A'}\varphi^\star_{A'A}$ is completely mixed on its support.
\end{IEEEproof}

\section{Proof of Proposition~\ref{prop:upper-bound}}\label{appx:proof-upper-bound}

\begin{IEEEproof}
That $\chi_\rho(\cN) - \chi(\cN) \leq
\Mutual_\rho(\cN)-\chi(\cN)$ follows trivially from Proposition~\ref{prop:chi-mutual-ea}.

We now show $\Mutual_\rho(\cN)-\chi(\cN)\leq D_F(\rho_{E_AE_B})$. Assume $D_F(\rho_{E_AE_B})$ is achieved by the
decomposition $\rho_{E_AE_B}=\sum_zq_Z(z)\rho^z_{E_AE_B}$ and for each $\rho^z_{E_AE_B}$, $D_R(\rho^z_{E_AE_B})$ is
achieved by the orthonormal bases $\{\vert\phi_{y\vert z}\rangle\}$ in system $E_B$. For each $y$ and $z$, define the
following conditional probability and state:
\begin{equation}
  p_{Y\vert Z}(y\vert z):=\tr\langle\phi_{y\vert z}\vert\rho_{E_AE_B}\vert\phi_{y\vert z}\rangle,\quad   
  \rho_{E_A}^{y\vert z}:=\langle\phi_{y\vert z}\vert\rho_{E_AE_B}\vert\phi_{y\vert z}\rangle/p_{Y\vert Z}(y\vert z).
\end{equation}
Then by assumption we have
\begin{IEEEeqnarray}{rCl}
  D_F(\rho_{E_AE_B})  &=& \sum_zp_Z(z)D_R\left(\rho_{E_AE_B}^z\right), \label{eq:D_F-1}\\
  D_R\left(\rho_{E_AE_B}^z\right) &=& \Rel\left(\rho_{E_AE_B}^z\rel
          \sum_yp_{Y\vert Z}(y\vert z)\rho_{E_A}^{y\vert z}\ox\proj{\phi_{y\vert z}}_Y\right),\; \forall z,
          \label{eq:D_R-1}
\end{IEEEeqnarray}
where we use classical symbol $Y$ to represent the collapsed quantum system $E_B$. Define the following
classical-quantum states:
\begin{IEEEeqnarray}{rCl}
  \rho_{ZE_AE_B} &:=& \sum_zp_Z(z)\proj{z}_Z\ox\rho^z_{E_AE_B}, \\
  \ol{\rho}_{ZE_AY}
&:=& \sum_{y,z}p_{Y,Z}(y, z)\proj{z}_Z\ox\rho_{E_A}^{y\vert z}\ox\proj{\phi_{y\vert z}}_{Y},
\end{IEEEeqnarray}
where $p_{Y,Z}(y, z):=p_{Y\vert Z}(y\vert z)p_Z(z)$. $\sigma_{ZE_AY}$ is obtained from $\rho_{ZE_AE_B}$ by performing
the conditional measurement $\{\vert\phi_{y\vert z}\rangle\}$ on each conditional state $\rho^z_{E_AE_B}$.

Assume $\{p_X,\cE^x_{E_A\to A}\}$ achieves $\Mutual_\rho(\cN)$. We define two new classical-quantum states by treating
$\rho_{ZE_AE_B}$ and $\ol{\rho}_{ZE_AY}$ as assistance states and $\{p_X,\cE^x_{E_A\to A}\}$ as encoding operations,
respectively:
\begin{IEEEeqnarray}{rCl}
  \sigma_{XZBE_B} 
&:=& \sum_xp_X(x)\proj{x}_X\ox\cN_{A\to B}\circ\cE^x_{E_A\to B}(\rho_{ZE_AE_B}) \\
&=& \sum_{x,z}p_X(x)p_Z(z)\proj{x}_X\ox\proj{z}_Z\ox\cN\circ\cE^x(\rho^z_{E_AE_B}), \\
  \ol{\sigma}_{XZBY}
&:=& \sum_xp_X(x)\proj{x}_X\ox\cN_{A\to B}\circ\cE^x_{E_A\to B}(\ol{\rho}_{ZE_AY}) \\
&=& \sum_{x,y,z} p_X(x)p_{Y,Z}(y, z)\proj{x}_X\ox\proj{z}_Z
      \ox\cN\circ\cE^x(\rho_{E_A}^{y\vert z})\ox\proj{\phi_{y\vert z}}_Y.
\end{IEEEeqnarray}
For $\sigma_{XZBE_B}$ and $\ol{\sigma}_{XZBY}$ defined above, we have the following reduced states:
\begin{IEEEeqnarray}{rCl}
  \sigma_{XBE_B} 
&=& \tr_Z\sigma_{XZBE_B} \\
&=& \sum_xp_X(x)\proj{x}_X\ox\cN\circ\cE^x\left(\sum_zp_Z(z)\rho^z_{E_AE_B}\right) \\
&=& \sum_xp_X(x)\proj{x}_X\ox\cN\circ\cE^x\left(\rho_{E_AE_B}\right), \\
\sigma_{XZE_B} 
&=& \tr_B\sigma_{XZBE_B} = \sum_{x,z}p_X(x)p_Z(z)\proj{x}_X\ox\proj{z}_Z\ox\rho^z_{E_B}, \\
\sigma_B 
&=& \tr_{XZE_B}\sigma_{XZBE_B} = \sum_xp_X(x)\cN\circ\cE^x\left(\rho_{E_A}\right), \\
\ol{\sigma}_{XZY} 
&=& \tr_B\ol{\sigma}_{XZBY} =  \sum_{x,z}p_X(x)p_{Y,Z}(y,z)
                                \proj{x}_X\ox\proj{z}_Z\ox\proj{\phi_{y\vert z}}_Y, \\
\ol{\sigma}_B 
&=& \tr_{XZY}\ol{\sigma}_{XZBY} = \sum_xp_X(x)\cN\circ\cE^x\left(\rho_{E_A}\right).
\end{IEEEeqnarray}
We have
\begin{equation}
  \Mutual_\rho(\cN) = \Mutual(XE_B{:}B)_\sigma \leq \Mutual(XZE_B{:}B)_\sigma,
\end{equation}
where the equality follows from assumption and the inequality follows from data processing inequality.
What's more, since $X$, $Y$, and $Z$ are all classical systems, we have
\begin{equation}
  \Mutual(XYZ{:}B)_{\ol{\sigma}} \leq \chi(\cN).
\end{equation}
Consider now the following chain of inequalities
\begin{IEEEeqnarray}{rCl}
&~&   \Mutual_\rho(\cN) - \chi(\cN) \\
&\leq& \Mutual(XZE_B{:}B)_\sigma - \Mutual(XYZ{:}B)_{\ol{\sigma}} \\
&=&  \Shannon(B)_\sigma + \Shannon(XZE_B)_\sigma - \Shannon(XZBE_B)_\sigma 
    - \Shannon(B)_{\ol{\sigma}} - \Shannon(XZY)_{\ol{\sigma}} + \Shannon(XZBY)_{\ol{\sigma}} \\
&=&  \left[\Shannon(XZE_B)_\sigma - \Shannon(XZY)_{\ol{\sigma}}\right]
    + \left[\Shannon(XZBY)_{\ol{\sigma}} - \Shannon(XZBE_B)_\sigma\right] \\
&\leq& \left[\Shannon(XZBY)_{\ol{\sigma}} - \Shannon(XZBE_B)_\sigma\right]\label{eq:bound-eq1} \\
&=& \Rel\left(\sigma_{XZBE_B}\rel\ol{\sigma}_{XZBE_B}\right)\label{eq:bound-eq2} \\
&=& \sum_{x,z}p_X(x)p_Z(z)\Rel\left(\cN\circ\cE^x(\rho^z_{E_AE_B})\rel
        \sum_yp_{Y\vert Z}(y\vert z)
        \cN\circ\cE^{x}(\rho_{E_A}^{y\vert z})\ox\proj{\phi_{y\vert z}}_Y\right)\label{eq:bound-eq3} \\
&\leq& \sum_zp_Z(z)\Rel\left(\rho^z_{E_AE_B}\rel
        \sum_yp_{Y\vert Z}(y\vert z)\rho_{E_A}^{y\vert z}\ox\proj{\phi_{y\vert z}}_Y\right)\label{eq:bound-eq4} \\
&=& \sum_zp_Z(z)D_R\left(\rho_{E_AE_B}^z\right)\label{eq:bound-eq5} \\
&=& D_F(\rho_{E_AE_B})\label{eq:bound-eq6},
\end{IEEEeqnarray}
where~\eqref{eq:bound-eq1} follows from the fact that projective measurement 
increases entropy~\cite[Theorem 11.9]{nielsen2011quantum}, 
\eqref{eq:bound-eq2} follows from 
that $\tr\ol{\sigma}_{XZBY}\log\ol{\sigma}_{XZBY}=\tr\sigma_{XZBE_B}\log\ol{\sigma}_{XZBY}$,
\eqref{eq:bound-eq3} follows from the direct-sum property,
\eqref{eq:bound-eq4} follows from the data processing inequality,
\eqref{eq:bound-eq5} follows from~\eqref{eq:D_R-1}, 
and~\eqref{eq:bound-eq6} follows from~\eqref{eq:D_F-1}.
We are done.
\end{IEEEproof}

\section{Semiproduct operations}\label{appx:semi-global}

Here we show that the semi-global operation definition in~\eqref{eq:semi-global-channel} does cover all operations
composed of local operations and permutations. In its most general form, an operation that is composed solely by local
operations and permutations can be viewed as many rounds of ``permutation followed by local operation'':
\begin{equation}\label{eq:sequential-1}
  \underbrace{\cP^{\pi_1} \rightarrow \bigotimes_{i}\cE^{[i]\vert 1}}_{\textit{$1$-th round}}
\quad\rightarrow\quad
  \underbrace{\cP^{\pi_2} \rightarrow \bigotimes_{i}\cE^{[i]\vert 2}}_{\textit{$2$-th round}}
\quad\rightarrow\quad
  \cdots
\quad\rightarrow\quad
  \underbrace{\cP^{\pi_n} \rightarrow \bigotimes_{i}\cE^{[i]\vert n}}_{\textit{$n$-th round}}
\end{equation}
where the $\rightarrow$ indicates the state evolution direction. Let's now go depth into the ``permutation followed by
local operation'' structure. We can actually exchange the sequential order of the permutation and the local operations
by performing first the local operations in the order dominated by $\pi^{-1}$ (the inverse of $\pi$) and then the
permutation operation without changing the output state. That is,
\begin{equation}
  \cP^{\pi_1} \rightarrow \bigotimes_{i}\cE^{[i]\vert 1}
\quad\equiv\quad
  \bigotimes_{i}\cE^{[\pi^{-1}_1(i)]\vert 1} \rightarrow \cP^{\pi_1} 
\end{equation}
Adopting this exchange approach to the first round in~\eqref{eq:sequential-1}, we get
\begin{equation}\label{eq:sequential-2}
  \underbrace{\bigotimes_{i}\cE^{[\pi^{-1}_1(i)]\vert 1}}_{\textit{$1$-th round}}
\quad\rightarrow\quad
  \underbrace{\cP^{\pi_2\pi_1} \rightarrow \bigotimes_{i}\cE^{[i]\vert 2}}_{\textit{$2$-th round}}
\quad\rightarrow\quad
  \cdots
\quad\rightarrow\quad
  \underbrace{\cP^{\pi_n} \rightarrow \bigotimes_{i}\cE^{[i]\vert n}}_{\textit{$n$-th round}}
\end{equation}
Repeating this approach $n$ times, the operation given in~\eqref{eq:sequential-1} becomes
\begin{equation}\label{eq:sequential-3}
  \bigotimes_{i}\cE^{[\pi^{-1}_1(i)]\vert 1}
\rightarrow
  \bigotimes_{i}\cE^{[(\pi_2\pi_1)^{-1}(i)]\vert 2}
\rightarrow
  \cdots
\rightarrow
  \bigotimes_{i}\cE^{[(\pi_n\cdots\pi_2\pi_1)^{-1}(i)]\vert n}
\rightarrow
  \cP^{\pi_n\cdots \pi_2\pi_1},
\end{equation}
which is exactly of the form given in~\eqref{eq:semi-global-channel}.

\section{Proof of Proposition~\ref{prop:pinching-bound}}\label{appx:prop:pinching-bound}

\begin{IEEEproof}
We will show the following strict inequality:
\begin{equation}\label{eq:pinching-bound-1}
  \chi_{\varphi^\star_{\bm{A}'\bm{A}}}(\cP_{\bm{A}\to \bm{A}}) 
< \Mutual_{\varphi^\star_{\bm{A}'\bm{A}}}(\cP_{\bm{A}\to \bm{A}}) = \Mutual(\cP_{\bm{A}\to \bm{A}}),
\end{equation}
which clearly implies Proposition~\ref{prop:pinching-bound}. To show the equality in~\eqref{eq:pinching-bound-1}, we
construct explicitly a state $\omega_{\bm{U}\bm{A}\bm{A}'}$ for which
$\Mutual(\bm{U}\bm{A}'{:}\bm{A})_\omega=\Mutual(\cP_{\bm{A}\to \bm{A}})$. The equality then follows from
Proposition~\ref{prop:mutual-relation}. To show the strict inequality in~\eqref{eq:pinching-bound-1}, we recall that
$\rho^\star_{\bm{A}}=\tr_{\bm{A}'}\varphi^\star_{\bm{A}'\bm{A}}$ is not completely mixed on its support by assumption.
This fact together with Proposition~\ref{prop:chi-mutual-equal-condition} yields the strict inequality.

Let $\{W_{u_i}\}$ be a complete set of Weyl operators of subspace $A_i$ and let
$\cW_{u_i}(\cdot):=W_{u_i}(\cdot)W_{u_i}^{\dagger}$ be the corresponding unitary channel. Consider the following set of
unitary channels on system $\bm{A}$:
\begin{equation}
  \left\{ \cW_{\bm{u}}: \bm{u}=(u_1,\cdots,u_k), u_i = 0,\cdots, d^2_i-1\right\},
\end{equation}
where $\cW_{\bm{u}}$ is understood as that $\cW_{u_i}$ is performed on subspace $A_i$. Note that these unitary channels
are commutative to the subspace projections $\Pi_i$. The size of this set is $D\equiv\prod_{i=1}^kd^2_i$. Let $\bm{U}$
be a $D$-dimensional classical system. Consider the following classical-quantum states:
\begin{IEEEeqnarray}{rCl}
  \omega^{\bm{u}}_{\bm{A}'\bm{A}} 
&:=& \cP_{\bm{A}\to\bm{A}}\circ\cW_{\bm{u}}\left(\varphi^\star_{\bm{A}'\bm{A}}\right), \\
  \omega_{\bm{U}\bm{A}'\bm{A}} 
&:=& \frac{1}{D}\sum_{\bm{u}}\proj{\bm{u}}_{\bm{U}}\ox\omega^{\bm{u}}_{\bm{A}'\bm{A}}.
\end{IEEEeqnarray}
The reduced state $\omega_{\bm{A}'\bm{A}}$ has the form 
\begin{equation}
    \omega_{\bm{A}'\bm{A}} 
=  \tr_{\bm{U}}\omega_{\bm{U}\bm{A}'\bm{A}}
=  \frac{1}{D}\sum_{\bm{u}}
    \cP_{\bm{A}\to\bm{A}}\circ\cW_{\bm{u}}\left(\varphi^\star_{\bm{A}'\bm{A}}\right)
=  \sum_{i=1}^kp^\star_i\pi_i\ox\pi_i,
\end{equation}
where $\pi_i$ is the completely mixed state of system $A_i$. Since $\bm{U}$ is classical,
$\omega_{\bm{U}\bm{A}'\bm{A}}$ forms a feasible solution to
$\Mutual_{\varphi^\star_{\bm{A}'\bm{A}}}(\cP_{\bm{A}\to\bm{A}})$. We have
\begin{IEEEeqnarray}{rCl}
  \Mutual_{\varphi^\star_{\bm{A}'\bm{A}}}(\cP_{\bm{A}\to \bm{A}})
&\geq& \Mutual(\bm{U}\bm{A}'{:}\bm{A})_\omega \\
&=& \Shannon(\bm{A}')_\omega + \Shannon(\bm{A})_\omega - \Shannon(\bm{A}'\bm{A}{\vert}\bm{U})_\omega \\
&=& 2\left(\Shannon(\bm{p}^\star) + \sum_{i=1}^kp^\star_i\log d_i\right) - \Shannon(\bm{p}^\star) \\
&=& \log\Delta,
\end{IEEEeqnarray}
where the last inequality follows from~\eqref{eq:pinching-1}. Since $\Mutual(\cP)=\log\Delta$, we conclude that
$\omega_{\bm{U}\bm{A}'\bm{A}}$ is an optimal state achieving
$\Mutual_{\varphi^\star_{\bm{A}'\bm{A}}}(\cP_{\bm{A}\to\bm{A}})$.

From the above argument, we easily obtain the following lower bound on
$\chi_{\varphi^\star_{\bm{A}'\bm{A}}}(\cP_{\bm{A}\to\bm{A}})$, since $\omega_{\bm{U}\bm{A}'\bm{A}}$ forms a feasible
solution:
\begin{IEEEeqnarray}{rCl}
  \chi_{\varphi^\star_{\bm{A}'\bm{A}}}(\cP_{\bm{A}\to \bm{A}})
&\geq& \Mutual(\bm{U}{:}\bm{A}'\bm{A})_\omega \\
&=& \Shannon(\bm{A}'\bm{A})_\omega - \Shannon(\bm{A}'\bm{A}{\vert}\bm{U})_\omega \\
&=& \Shannon(\bm{p}^\star) + 2\sum_{i=1}^kp^\star_i\log d_i - \Shannon(\bm{p}^\star) \\
&=& \log\Delta - \Shannon(\bm{p}^\star).
\end{IEEEeqnarray}
\end{IEEEproof}

\section{Proof of Proposition~\ref{prop:covariant}}\label{appx:prop:covariant}

\begin{IEEEproof}
Assume $\cE^\ast_{E_A \to A}$ achieves $\min_{\cE_{E_A \to A}} \Shannon(\cN_{A\to B}\circ\cE_{E_A \to
A}(\rho_{E_AE_B}))$ and let
\begin{equation}
    \Xi\equiv\Shannon(\cN_{A\to B}(\pi_A)) + \Shannon(\rho_{E_B}) 
- \Shannon(\cN_{A\to B}\circ\cE^\ast_{E_A \to A}(\rho_{E_AE_B})).
\end{equation}
We will show the following chain of inequalities which trivially implies~\eqref{eq:covariant}:
\begin{equation}\label{eq:covariant-2}
    \Xi \leq \chi_\rho(\cN) \leq \Mutual_\rho(\cN) \leq \Xi.
\end{equation} 

To show the first inequality of~\eqref{eq:covariant-2}, we choose the following encoding operations $\cE^g(\cdot)
:= f_A(g)\cE^\ast(\cdot)f_A(g)^\dagger$, where $g$ is subject to the Haar measure $\mu$. The corresponding
classical-quantum state is
\begin{equation}
  \omega_{GBE_B} := \int_G\mu(dg) \proj{g}_G \ox \cN_{A\to B}\circ\cE^g(\rho_{E_AE_B}).
\end{equation}
One can check $\omega_{BE_B}=\cN_{A\to B}(\pi_A) \ox \rho_{E_B}$.
By the definition of $\chi_\rho(\cN)$, it holds that
\begin{IEEEeqnarray}{rCl}
  \chi_\rho(\cN) 
&\geq& \Mutual(G{:}BE_B)_\omega \\
&=& \Shannon(BE_B)_\omega - \Shannon(BE_B{\vert}G)_\omega \\
&=& \Shannon\left(\cN_{A\to B}(\pi_A)\right) + \Shannon(\rho_{E_B}) 
  - \int_G\mu(dg) \Shannon\left(\cN_{A\to B}\circ\cE^g(\rho_{E_AE_B})\right) \\
&=& \Shannon\left(\cN_{A\to B}(\pi_A)\right) + \Shannon(\rho_{E_B})
  - \int_G\mu(dg)\Shannon\left(\cN_{A\to B}\left(f_A(g)\cE^\ast(\rho_{E_AE_B})f_A(g)^\dagger\right)\right)
    \label{eq:covariant-3} \\
&=& \Shannon\left(\cN_{A\to B}(\pi_A)\right) + \Shannon(\rho_{E_B})
  - \int_G\mu(dg)\Shannon\left(f_B(g)\cN_{A\to B}\circ\cE^\ast(\rho_{E_AE_B})f_B(g)^\dagger\right)
    \label{eq:covariant-4} \\
&=& \Shannon\left(\cN_{A\to B}(\pi_A)\right) + \Shannon(\rho_{E_B})
  - \Shannon\left(\cN_{A\to B}\circ\cE^\ast(\rho_{E_AE_B})\right) \\
&\equiv& \Xi,  
\end{IEEEeqnarray}
where~\eqref{eq:covariant-3} follows from the definition of $\cE^g$ 
and~\eqref{eq:covariant-4} follows since $\cN$ is covariant.

The second inequality of~\eqref{eq:covariant-2} was proved in Proposition~\ref{prop:chi-mutual-ea}.

To show the third inequality of~\eqref{eq:covariant-2}, assume $\omega_{XBE_B}$ 
defined in~\eqref{eq:omega_XBEB} achieves $\Mutual_\rho(\cN)$. Then
\begin{IEEEeqnarray}{rCl}
  \Mutual_\rho(\cN) 
= \Mutual(XE_B{:}B)_\omega
&=& \Shannon(B)_\omega + \Shannon(E_B)_\omega - \Shannon(BE_B{\vert}X)_\omega \\
&=& \Shannon(B)_\omega + \Shannon(E_B)_\rho - \sum_xp_X(x)\Shannon(\cN\circ\cE^x(\rho_{E_AE_B})) \\
&\leq& \Shannon(B)_\omega + \Shannon(E_B)_\rho - \Shannon(\cN_{A\to B}\circ\cE^\ast_{E_A \to A}(\rho_{E_AE_B})),
    \label{eq:covariant-8}
\end{IEEEeqnarray}
where the inequality follows from the assumption of $\cE^\ast_{E_A \to A}$.
On the other hand, it holds that
\begin{IEEEeqnarray}{rCl}
  \Shannon(B)_\omega
&=& \Shannon\left(\sum_xp_X(x)\cN\circ\cE^x(\rho_{E_A})\right) \\
&=& \int_G\mu(dg)\Shannon\left(f_B(g)\left(\sum_xp_X(x)\cN\circ\cE^x(\rho_{E_A})\right)f_B(g)^\dagger\right) \\
&=& \int_G\mu(dg)\Shannon\left(\sum_xp_X(x)\cN\left(f_A(g)\cE^x(\rho_{E_A})f_A(g)^\dagger\right)\right)
    \label{eq:covariant-5} \\
&\leq& \Shannon\left(\sum_xp_X(x)\cN\left(\int_G\mu(dg) f_A(g)\cE^x(\rho_{E_A})f_A(g)^\dagger\right)\right)
    \label{eq:covariant-6} \\
&=& \Shannon\left(\cN(\pi_A)\right),\label{eq:covariant-7}
\end{IEEEeqnarray}
where~\eqref{eq:covariant-5} follows since $\cN$ is covariant and~\eqref{eq:covariant-6} follows from the
concavity of quantum entropy. Combining~\eqref{eq:covariant-8} and~\eqref{eq:covariant-7}, we get
$\Mutual_\rho(\cN)\leq\Delta$.
\end{IEEEproof}

\section{Proof of Proposition~\ref{prop:erasure-channel}}\label{appx:prop:erasure-channel}

\begin{IEEEproof}
Notice that $\Shannon(\cE_p(\pi)) = (1-p)\log d + \Binary(p)$ and
$\Shannon(E_B)_{\Phi_{\bm{\lambda}}}=\Shannon(\bm{\lambda})$, by Theorem~\ref{thm:capacity} and
Proposition~\ref{prop:covariant} it is equivalent to show that
\begin{equation}\label{eq:erasure-equality}
  \min_{\cE} \Shannon(\cE_p\circ\cE(\Phi_{\bm{\lambda}}))
= \Binary(p) + p\Shannon(\bm{\lambda}).
\end{equation}
Let $\sigma_{BE_B} := \cE_p\circ\cE(\Phi_{\bm{\lambda}}) 
        = (1-p)\cE(\Phi_{\bm{\lambda}}) + p\proj{e}\ox\Phi_{\bm{\lambda}}^{E_B}$.
Consider the following isometry:
\begin{equation}
    U_{B\to BY} := \Pi_B\ox\ket{0}_Y + \proj{e}_B\ox\ket{1}_Y
\end{equation}
and the corresponding induced state
\begin{equation}
  \omega_{YBE_B} 
:= U_{B\to BY} \sigma_{BE_B} U_{B\to BY}^\dagger 
= (1-p)\cE^x(\Phi_{\bm{\lambda}})\ox\proj{0}_Y 
                                  + p\proj{e}_B\ox\Phi_{\bm{\lambda}}^{E_B}\ox\proj{1}_Y.
\end{equation}
We have the following chain of inequalities:
\begin{IEEEeqnarray}{rCl}
   \Shannon(\cE_p\circ\cE(\Phi_{\bm{\lambda}}))
&=& \Shannon(BE_B)_\sigma \\
&=& \Shannon(BE_BY)_\omega\label{eq:erasure-1} \\
&=& \Shannon(Y)_\omega + \Shannon(BE_B\vert Y)_\omega \\
&=& \Binary(p) + (1-p)\Shannon(\cE(\Phi_{\bm{\lambda}}))
  + p\Shannon\left(\proj{e}_B\ox\Phi_{\bm{\lambda}}^{E_B}\right) \\
&=& \Binary(p) + p\Shannon(\bm{\lambda}) + (1-p)\Shannon(\cE(\Phi_{\bm{\lambda}})) \\
&\geq& \Binary(p) + p\Shannon(\bm{\lambda}),\label{eq:erasure-2}
\end{IEEEeqnarray}
where~\eqref{eq:erasure-1} follows since quantum entropy is isometry invariant and~\eqref{eq:erasure-2}
follows as the entropy is non-negative. Furthermore, the equality in~\eqref{eq:erasure-2} is attainable by
choosing $\cE$ to be the identity channel. This concludes~\eqref{eq:erasure-equality}.
\end{IEEEproof}

\end{document}